\documentclass[10pt,twocolumn,twoside]{IEEEtran}

\usepackage{amsmath}
\usepackage{amssymb}
\usepackage{latexsym}
\usepackage{verbatim}
\usepackage{subfigure}
\usepackage[final]{graphicx}
\usepackage{psfrag}

\newtheorem{theorem}{Theorem}
\newtheorem{proposition}{Proposition}
\newtheorem{lemma}{Lemma}

\newtheorem{remark}{Remark}

\newtheorem{example}{Example}

{\begin{list}               %    with flush left bullets.
    {$\bullet$ \hfill}{
        \setlength{\leftmargin}{\parindent}
        \setlength{\parsep}{0.04\baselineskip}
        \setlength{\itemsep}{0.5\parsep}
        \setlength{\labelwidth}{\leftmargin}
        \setlength{\labelsep}{0em}}
    }
{\end{list}}

\providecommand{\eref}[1]{\eqref{eq:#1}}  % call \eqref from amstex
\providecommand{\cref}[1]{Chapter~\ref{chap:#1}}
\providecommand{\sref}[1]{Section~\ref{sec:#1}}
\providecommand{\fref}[1]{Figure~\ref{fig:#1}}

\providecommand{\R}{\ensuremath{\mathbb{R}}}
\providecommand{\C}{\ensuremath{\mathbb{C}}}

\providecommand{\abs}[1]{\lvert#1\rvert}
\providecommand{\norm}[1]{\lVert#1\rVert}

\providecommand{\set}[1]{\left\{#1\right\}}

\providecommand{\bydef}{\overset{\text{def}}{=}}

\renewcommand{\vec}[1]{\ensuremath{\boldsymbol{#1}}}
\providecommand{\mat}[1]{\ensuremath{\boldsymbol{#1}}}

\providecommand{\mA}{\mat{A}} 
 
\providecommand{\mD}{\mat{D}}

\providecommand{\mI}{\mat{I}}  
  
\providecommand{\mM}{\mat{M}} \providecommand{\mP}{\mat{P}} 
 
\providecommand{\mS}{\mat{S}} \providecommand{\mU}{\mat{U}} 

\providecommand{\mW}{\mat{W}}

\providecommand{\mZ}{\mat{Z}}

\providecommand{\va}{\vec{a}} 
 \providecommand{\ve}{\vec{e}}

\providecommand{\vq}{\vec{q}} \providecommand{\vs}{\vec{s}}

\providecommand{\vu}{\vec{u}} \providecommand{\vw}{\vec{w}}
\providecommand{\vx}{\vec{x}} \providecommand{\vy}{\vec{y}}
 
 \providecommand{\vzero}{\vec{0}}
 \providecommand{\vv}{\vec{v}}

\usepackage{cite}

\usepackage{color}

\usepackage{booktabs,multirow}

\usepackage{enumitem}
\usepackage{pmat}
\usepackage{commath}

\usepackage{algorithm, algorithmic}

\usepackage{hyperref}

\usepackage{esdiff}

\usepackage{pgf,tikz,psfrag}

\usepackage{dsfont}
\newcommand{\charfn}{\mathds{1}}

\newcommand{\argmin}{\operatornamewithlimits{argmin}}
\newcommand{\asc}{\overset{\text{a.s.}}{\longrightarrow}}
\newcommand{\cip}{\overset{\mathcal{P}}{\longrightarrow}}

\providecommand{\lm}{\lambda}
\providecommand{\blm}{\overline{\lambda}}

\providecommand{\slma}{\lambda^\ast_\alpha}
\providecommand{\muf}{\phi}
\providecommand{\uf}{\psi}
\providecommand{\lf}{\zeta}
\providecommand{\ub}{\tau}
\providecommand{\nrm}{\kappa}

\DeclareMathOperator{\diag}{diag}
\DeclareMathOperator{\adj}{adj}
\DeclareMathOperator{\sign}{sign}
\DeclareMathOperator{\Tr}{Tr}

\providecommand{\vxi}{\vec{\xi}}

\begin{document}
%.
% Paper Title
\title{Phase Transitions of Spectral Initialization for High-Dimensional Nonconvex Estimation}

\author{%
Yue M. Lu and Gen Li
\thanks{Y. M. Lu is with the John A. Paulson School of Engineering and Applied Sciences, Harvard University, Cambridge, MA 02138, USA (e-mail:  \href{mailto:yuelu@seas.harvard.edu}{yuelu@seas.harvard.edu}). Part of this work was done during his visit to the Information Initiative at Duke (iiD) in Spring 2016. He thanks members of this interdisciplinary program for their hospitality.}%
\thanks{G. Li is with the Department of Electronic Engineering, Tsinghua University, Beijing 100084, China (e-mail: \href{mailto:g-li16@mails.tsinghua.edu.cn}{g-li16@mails.tsinghua.edu.cn}). He was a summer visiting undergraduate student at the John A. Paulson School of Engineering and Applied Sciences, Harvard University.}%
\thanks{This work was supported in part by the ARO under contract W911NF-16-1-0265 and by the US National Science Foundation under grants CCF-1319140 and CCF-1718698. Preliminary version of this work was presented at the IEEE International Symposium on Information Theory (ISIT) in 2017.}
}

\markboth{} {Lu \MakeLowercase{and} Li: Spectral Initialization for High-Dimensional Nonconvex Estimation}

\maketitle

\begin{abstract}
We study a spectral initialization method that serves a key role in recent work on estimating signals in nonconvex settings. Previous analysis of this method focuses on the phase retrieval problem and provides only performance bounds. In this paper, we consider arbitrary generalized linear sensing models and present a precise asymptotic characterization of the performance of the method in the high-dimensional limit. Our analysis also reveals a phase transition phenomenon that depends on the ratio between the number of samples and the signal dimension. When the ratio is below a minimum threshold, the estimates given by the spectral method are no better than random guesses drawn from a uniform distribution on the hypersphere, thus carrying no information; above a maximum threshold, the estimates become increasingly aligned with the target signal. The computational complexity of the method, as measured by the spectral gap, is also markedly different in the two phases. Worked examples and numerical results are provided to illustrate and verify the analytical predictions. In particular, simulations show that our asymptotic formulas provide accurate predictions for the actual performance of the spectral method even at moderate signal dimensions.
\end{abstract}


\section{Introduction}
\label{sec:intro}

We consider the problem of estimating an $n$-dimensional vector $\vxi$ from a number of generalized linear measurements. Let $\set{\va_i}_{1 \le i \le m}$ be a set of sensing vectors in $\R^n$. Given $\set{\va_i^\top \vxi}$, the measurements are drawn independently from 
\begin{equation}\label{eq:model}
y_i \sim f(y \, \vert \, \va_i^\top \vxi),
\end{equation}
where $f(\cdot \, \vert \, \cdot)$ is a conditional density function modeling the acquisition process. This model arises in many problems in signal processing and statistical learning. Examples include photon-limited imaging \cite{UnserE:88, Yang:2012vn}, phase retrieval \cite{Fienup:82}, signal recovery from quantized measurements \cite{Rangan:2001uq}, and various single-index and generalized linear regression problems \cite{Dobson:2008, Hastie:2011qy}.

%We assume that $\vxi$ first goes through a linear transform associated with a set of sensing vectors $\set{\va_i}_{1 \le i \le m}$. The transform coefficients $\set{\va_i^\top \vxi}$ are then observed over a memoryless channel. Given $\vxi$, each measurement $y_i$, for $1 \le i \le m$, is drawn independently from

%Let $\set{\va_i}_{1 \le i \le m}$ be a set of sensing vectors in $\R^n$. We are interested in estimating an unknown vector $\vxi \in \R^n$ from a number of generalized linear measurements $\set{y_i}_{1 \le i \le m}$: given $\vxi$ and $\set{\va_i}$, the $i$th measurement $y_i$ is drawn independently from some distribution
%\begin{equation}\label{eq:model}
%\mathbb{P}_i(y) = f(y \, \vert \, \va_i^\top \vxi),
%\end{equation}
%where $f(\cdot \, \vert \, \cdot)$ is a conditional density function modeling the acquisition process. This model arises in many problems in signal processing and learning. Examples include photon-limited imaging \cite{UnserE:88, Yang:2012vn}, phase retrieval \cite{Fienup:82}, signal recovery from quantized measurements \cite{Rangan:2001uq}, and generalized linear regression \cite{Hastie:2011qy}.

The standard method for recovering $\vxi$ is to use the estimator
\begin{equation}\label{eq:regression}
\widehat{\vxi} = \underset{\vx}{\text{arg}\,\min} \sum_{i=1}^m \ell(y_i, \va_i^\top \vx),
\end{equation}
where $\ell: \R^2 \rightarrow \R$ is some loss function (\emph{e.g.}, the negative log-likelihood of the observation model as used in maximum likelihood estimation). In many applications, however, the natural loss function is not convex with respect to $\vx$. There is often no effective way to convexify \eref{regression}. In those cases for which convex relaxations do exist, the resulting algorithms can be computationally expensive. The problem of phase retrieval, where $y_i = (\va_i^\top \vxi)^2 + \varepsilon_i$ for some noise terms $\set{\varepsilon_i}$, is an example in the latter scenario. Convex relaxation schemes such as those based on lifting and semidefinite programming (\emph{e.g.}, \cite{Candes:2013xy, Candes:2014ty, Jaganathan:2013zl, Waldspurger:2015rz}) have been successfully developed for solving the phase retrieval problem, but the challenges facing these schemes lie in their actual implementation. In practice, the computational complexity and memory requirement associated with these convex-relaxation methods are prohibitive for signal dimensions that are encountered in real-word applications such as imaging.

In light of these issues, there is strong recent interest in developing and analyzing efficient iterative methods that directly solve nonconvex forms of \eref{regression}. Examples include the alternating minimization scheme for phase retrieval \cite{Netrapalli:2013qv}, the Wirtinger Flow algorithm and its variants \cite{Candes:2015fv, Chen:2015eu, Zhang:2016gf, WangGY:2016, Soltanolkotabi:2017}, iterative projection methods \cite{LiGL:15, ChiL:16}, and recent schemes for phase retrieval using linear programming \cite{Goldstein:2016, Bahmani:2016, DhifallahL:17, DhifallahTY:18}. A common ingredient that contributes to the success of these algorithms for nonconvex estimation is that they all use some carefully-designed spectral method as an initialization step, which is then followed by further (iterative) refinement. Beyond the signal estimation problem considered in this paper, related spectral methods have also been successfully applied to initialize algorithms for solving other nonconvex problems such as matrix completion \cite{KeshavanMO:10}, low-rank matrix recovery \cite{JainNS:13}, blind deconvolution \cite{LeeLJB:17, LiLSW:16}, sparse coding \cite{Arora:2015}, and joint alignment from pairwise differences \cite{ChenC:16}.

In this paper, we present an \emph{exact} high-dimensional analysis of a widely-used spectral method \cite{Netrapalli:2013qv, Candes:2015fv, Chen:2015eu} for estimating $\vxi$. The method consists of only two steps: First, construct a data matrix from the sensing vectors and measurements as
\begin{equation}\label{eq:D_mtx}
\mD_m \bydef \frac{1}{m} \sum_{i=1}^m \mathcal{T}(y_i) \va_i \va_i^\top,
\end{equation}
where $\mathcal{T}: \R \rightarrow \R$ is a preprocessing function (\emph{e.g.}, a trimming or truncation step). Second, compute a normalized eigenvector, denoted by $\vx_1$, that corresponds to the largest eigenvalue of $\mD_m$. The vector $\vx_1$ is then our estimate of $\vxi$ (up to an unknown scalar). It is notable that this method is \emph{model-free} in that the algorithm does not require the knowledge of the exact acquisition process [\emph{i.e.}, the conditional density $f(\cdot \, \vert \, \cdot)$ in \eref{model}].

The idea of this spectral method can be traced back to the early work of Li \cite{Li:92}, under the name of Principal Hessian Directions for general multi-index models. Similar spectral techniques were also proposed in \cite{KeshavanMO:10, JainNS:13}, for initializing algorithms for matrix completion. In \cite{Netrapalli:2013qv}, Netrapalli, Jain, and Sanghavi used this method to address the problem of phase retrieval. Under the assumption that the sensing vectors consist of i.i.d. Gaussian random variables, these authors show that the leading eigenvector $\vx_1$ is aligned with the target vector $\vxi$ in direction when there are sufficiently many measurements. More specifically, they show that the squared \emph{cosine similarity}
\begin{equation}\label{eq:rho}
\rho(\vxi, \vx_1) \bydef \frac{({\vxi^\top \vx_1})^2}{\norm{\vxi}^2\norm{\vx_1}^2},
\end{equation}
which measures the degree of the alignment between the two vectors, approaches $1$ with high probability, when the number of samples $m \ge c_1  n \log^3 n$. This sufficient condition on sample complexity was later improved to $m \ge c_2 n \log n$ in \cite{Candes:2013xy}, and further improved to $m \ge c_3 n$ in \cite{Chen:2015eu} with an additional trimming step on the measurements. In these expressions, $c_1, c_2, c_3$ stand for some unspecified numerical constants.

In this paper, we provide a precise asymptotic characterization of the performance of the spectral method under Gaussian measurements. Our analysis considers general acquisition models under arbitrary conditional distributions $f(y \,\vert\, \va_i^\top \vxi)$, of which the phase retrieval problem is a special case. Unlike previous work, which only provides bounds for $\rho(\vxi, \vx_1)$, we derive the exact high-dimensional limit of this value. In particular, we show that, as $n$ and $m$ both tend to infinity with the \emph{sampling ratio} $\alpha \bydef m/n$ kept fixed, the squared cosine similarity $\rho$ converges in probability to a limit value $\rho(\alpha)$. Explicit formulas are provided for computing $\rho(\alpha)$. 

\begin{figure}[t]
	\centering    
	\includegraphics[width=0.47\linewidth]{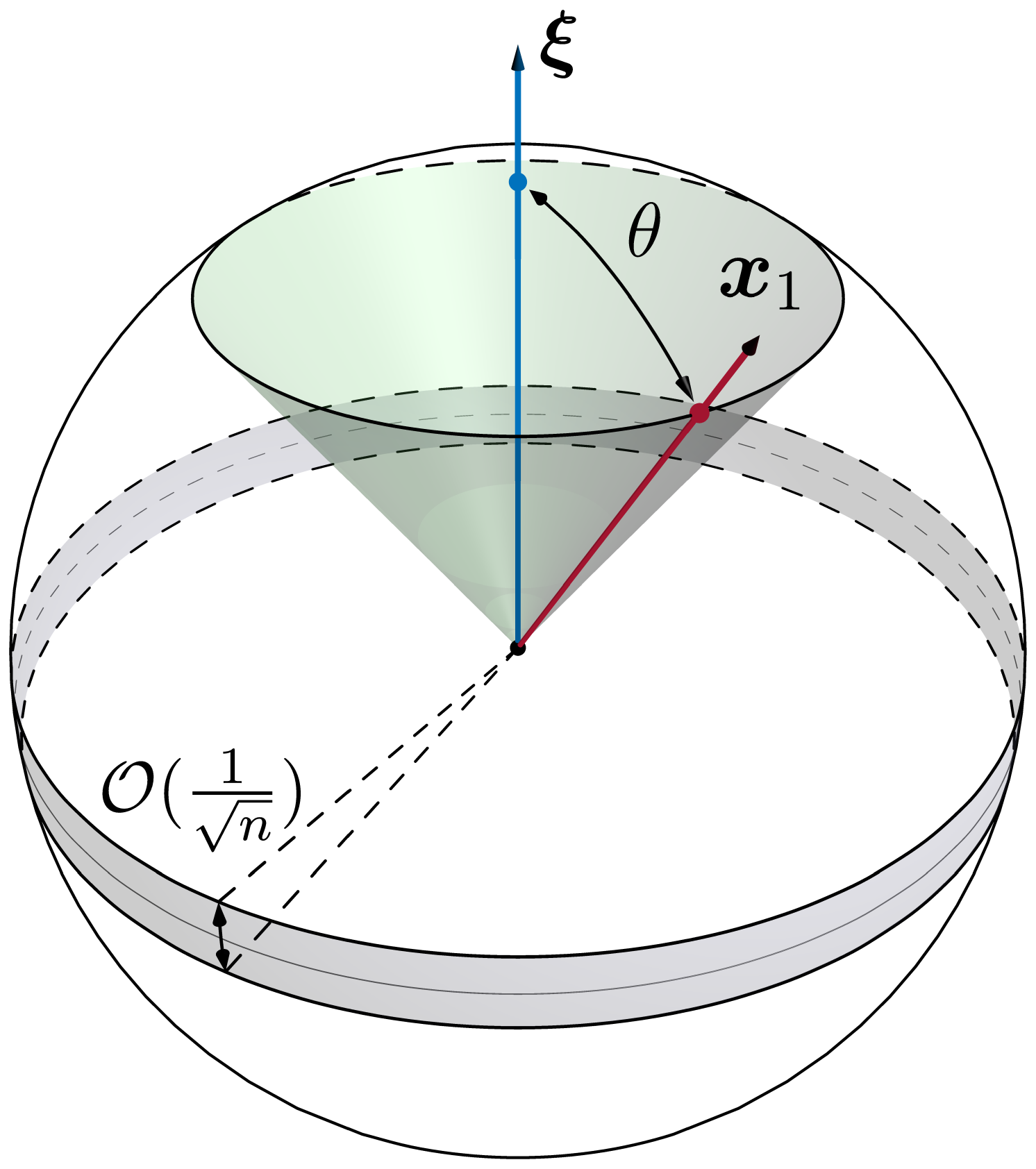}
	\caption{Illustrations of the phase transitions of the spectral method. Depending on the sampling ratio $\alpha$, the asymptotic performance of the spectral method can be in one of two very different phases: In the uncorrelated phase, its estimate $\vx_1$ is asymptotically orthogonal to $\vxi$. The performance in this case is no better than an arbitrary guess drawn uniformly at random from the hypersphere $\mathcal{S}^{n-1}$. In the correlated phase, the estimate $\vx_1$ (or its negative version $-\vx_1$) will be concentrated on the surface of a right circular cone, making an angle $\theta = \arccos\big(\sqrt{\rho(\alpha)}\big)$ with the target vector $\vxi$.}
	\label{fig:sphere}
\end{figure}

Geometrically, the squared cosine similarity $\rho(\vxi, \vx_1)$ as defined in \eref{rho} specifies the angle $\theta$ between $\vxi$ and $\vx_1$. The values of $\rho$ vary from $0$ to $1$: $\rho = 1$ means perfect alignment, \emph{i.e.}, $\theta = 0$ or $\pi$; and $\rho =  0$ is the opposite case, meaning $\vx_1$ is orthogonal to (\emph{i.e.} uncorrelated with) $\vxi$. That the spectral method can yield an estimate $\vx_1$ with a positive $\rho$ in high dimensional settings is a nontrivial property. To see this, assume that $\vxi$ is pointing towards the ``north pole'' in the unit $(n-1)$-sphere $\mathcal{S}^{n-1}$, as illustrated in \fref{sphere}. If we choose $\vx_1$ uniformly at random from $\mathcal{S}^{n-1}$, then with high probability, the resulting correlation $\sqrt{\rho(\vxi, \vx_1)}$ will be of order $\mathcal{O}(1/\sqrt{n})$. In other words, for large $n$, most of the uniform measure on $\mathcal{S}^{n-1}$ is concentrated within a very thin band of width $\mathcal{O}(1/\sqrt{n})$ near the ``equator'' of the sphere (see \fref{sphere}). 

%This peculiar \emph{concentration of measure} phenomenon in high dimensional geometry \cite{} makes it all the more remarkable that the spectral method, with sufficient measurements, is able to generate an estimate with a positive $\rho$. 

Our analysis reveals a phase transition phenomenon that occurs at certain critical values of the sampling ratio. In particular, there exist a lower and an upper threshold, denoted by $\alpha_{c,\min}$ and $\alpha_{c,\max}$, respectively, that mark the transitions between two very different phases.

(a) An \emph{uncorrelated phase} takes place when the sampling ratio $\alpha <  \alpha_{c,\min}$. Within this phase, the limiting value $\rho(\alpha) = 0$, meaning that the estimate from the spectral method is asymptotically uncorrelated with the target vector $\vxi$. In this case, the spectral method is not effective, as its estimate $\vx_1$ is no better than a random guess drawn uniformly from the hypersphere $\mathcal{S}^{n-1}$.

(b) A \emph{correlated phase} takes place when $\alpha > \alpha_{c,\max}$, with $\alpha_{c,\max}$ being the upper threshold. Within this phase, the limiting value $\rho(\alpha) > 0$. Geometrically, the estimate $\vx_1$ (or its negative version $-\vx_1$) will be concentrated on the surface of a right-circular cone (see \fref{sphere}) whose generating lines make an angle $\theta = \arccos\big(\sqrt{\rho(\alpha)}\big)$ to the target vector $\vxi$. Moreover, $\rho(\alpha)$ tends to 1 as $\alpha \rightarrow \infty$.

In many signal estimation models that we have studied so far, the two thresholds coincide, \emph{i.e.} $\alpha_{c,\min} = \alpha_{c,\max}$, meaning that the phase transition happens at a single critical value of the sampling ratio. However, it is indeed possible that $\alpha_{c,\min} < \alpha_{c,\max}$, in which case a finite number of correlated and uncorrelated phases alternative when $\alpha$ varies within the interval $(\alpha_{c,\min}, \alpha_{c,\max})$. A concrete example demonstrating this more complicated situation can be found in \sref{counter_example}.

The above phase transition phenomenon also has implications in terms of the computational complexity of the spectral method. In a correlated phase, there is a nonzero gap between the largest and the second largest eigenvalues of $\mD_m$. As a result, the leading eigenvector $\vx_1$ can be efficiently computed by using power iterations on $\mD_m$. In contrast, within an uncorrelated phase, the gap of the eigenvalues converges to zero, making power iterations inefficient.

The rest of the paper is organized as follows. After precisely laying out the various technical assumptions, we present in \sref{results} the main results of this work, stated as Theorem~\ref{thm:cos2} and Proposition~\ref{prop:alpha_c}. Examples and numerical simulations are also provided there to demonstrate and verify these analytical results. In particular, as a worked example, we derive a universal closed-form expression for the limiting values $\rho(\alpha)$ for all acquisition models that generate one-bit $\set{0, 1}$ measurements. We prove Theorem~\ref{thm:cos2} in \sref{proof}. Key to our proof is a deterministic, fixed-point characterization of the squared cosine similarity $\rho(\vxi, \vx_1)$, which is valid for any finite dimension $n$ and for any \emph{deterministic} sensing vectors $\set{\va_i}$. When specialized to Gaussian measurements, this fixed-point characterization allows us to connect our problem to a generalized version of the spiked population model (see, \emph{e.g.}, \cite{Johnstone:2001, Baik:2005, BaiY:12}) studied in random matrix theory. In \sref{phase}, we look more closely at the phase transition phenomenon predicted by our asymptotic results and prove Proposition~\ref{prop:alpha_c}. \sref{conclusion} concludes the paper with discussions on possible generalizations and improvements of our results as well as their connections to related work in the literature.

\emph{Notations}: To study the high-dimensional limit of the spectral initialization method, we shall consider a sequence of problem instances, indexed by the ambient dimension $n$. For each $n$, we seek to estimate an underlying signal denoted by $\vxi_n \in \R^n$. Formally, we should use $D_{m(n)}$ to denote the data matrix, where $m(n)$ is the number of measurements as a function of the dimension $n$. However, to lighten the notation, we will simply write it as $\mD_m$, keeping the dependence of $m$ on $n$ implicit. $\vx_1^n$ stands for a leading eigenvector of $\mD_m$. We use $\cip$ and $\asc$ to denote convergence in probability and almost sure convergence, respectively. Let $\mM$ be a symmetric matrix. Its eigenvalues in descending order are written as $\lambda_1^{\mM} \ge \lambda_2^{\mM} \ge \ldots \ge \lambda_n^{\mM}$. In particular, $\lambda_1^{\mM}$, sometimes also written as $\lambda_1(\mM)$, denotes the largest eigenvalue of $\mM$. Throughout the paper, $\mA^{1/2}$ stands for the principal square root of a positive semi-definite symmetric matrix $\mA$. For any $a, b \in \R$, we write $\max\set{a, b}$ as $a \vee b$. Finally, $\charfn_{x \in \mathcal{I}}$ stands for the indicator function of a set $\mathcal{I}$.

\section{Main Results}
\label{sec:results}

\subsection{Technical Assumptions}
\label{sec:assumptions}

In what follows, we first state the basic assumptions under which our results are proved.

\begin{enumerate}[label={(A.\arabic*)}]
\item \label{a:Gaussian} The sensing vectors are independent Gaussian random vectors. Specifically, let $(a_{ij})$, for $i, j \ge 1$, be a doubly infinite array of i.i.d. standard normal random variables. Then the $i$th sensing vector $\va_i = [a_{i1}, a_{i2}, \ldots, a_{in}]^\top$.

\item $m = m(n)$ with $\alpha_n = m(n) / n \rightarrow \alpha > 0$ as $n \rightarrow \infty$.

\item \label{a:norm} $\norm{\vxi_n} = \nrm > 0$.

\item \label{a:rvs} Let $s, y$ and $z$ be three random variables such that
\begin{equation}\label{eq:sy}
s \sim \mathcal{N}(0, 1),\ \mathbb{P}(y \,\vert\, s) = f(y \, \vert \, \nrm s), \ \text{and}\ z = \mathcal{T}(y),
\end{equation}
where $f(\cdot \,\vert\, \cdot)$ is the conditional density function \eref{model} associated with the observation model, and $\mathcal{T}(\cdot)$ is the preprocessing step used in the construction of $\mD_m$ in \eref{D_mtx}. We shall assume that the probability measure of the random variable $z$ is supported within a finite interval $[0, \tau]$. Throughout the paper, we always take $\tau$ to be the \emph{tightest} such upper bound.

\item \label{a:infty} As $\lm$ approaches $\tau$ from the right,
\begin{equation}\label{eq:ub_infty}
\lim_{\lm \rightarrow \tau^+} \mathbb{E}\frac{z}{(\lm - z)^2} = \lim_{\lm \rightarrow \tau^+} \mathbb{E}\frac{z s^2}{\lm - z} = \infty.
\end{equation}

\item \label{a:pos_corr} The random variables $z$ and $s^2$ are positively correlated: $\text{cov}(z, s^2) = \mathbb{E}\,z s^2 - \mathbb{E}z \, \mathbb{E}s^2  > 0$, which is equivalent to
\begin{equation}\label{eq:pos_corr}
\mathbb{E}\,z s^2 > \mathbb{E}z.
\end{equation}
\end{enumerate}

The last three assumptions require some explanations. First, we note that assumption~\ref{a:rvs} requires that $z$ should take values within a finite interval on the positive axis. This can be enforced by choosing a suitable function $\mathcal{T}(\cdot)$. For example, in the problem of phase retrieval, the measurement model ($y = s^2$) leads to unbounded $\set{y_i}$. We can set 
\begin{equation}\label{eq:trimming}
z = \mathcal{T}(y) = y \, \charfn_{\abs{y} \le t^2},
\end{equation}
where $t > 0$ is some parameter and $\charfn_{\abs{y} \le t^2}$ denotes the indicator function for the condition $\abs{y} \le t^2$. This is indeed the trimming strategy proposed in \cite{Chen:2015eu}. As shown there, this boundedness condition on the support of $z$ is an essential ingredient in achieving linear sample complexities. The assumption that $z$ be nonnegative is largely made to simplify our analysis, but this restriction can be removed. In a recent work \cite{MondelliM:17}, Mondelli and Montanari extended our results by showing that the same asymptotic predictions presented in this paper still hold under cases where $z$ can take negative values. See also Remark~\ref{rem:extension} in \sref{main_results}.

In assumption~\ref{a:infty}, the expressions in \eref{ub_infty} essentially require that the random variable $z$ should have sufficient probability mass near the upper bound $\ub$. Let $h(z) = \mathbb{E}_{s \vert z}(s^2 \vert z)$. We show in Appendix~\ref{appendix:infty} that \eref{ub_infty} holds when there exist some positive constants $c_0$ and $\varepsilon$ such that the probability density function $p_Z(z)$ of $z$ and the conditional moment $h(z)$ are both bounded below by $c_0$ for all $z \in [\ub - \varepsilon, \ub]$. The model in \eref{trimming} represents one such case. Another sufficient condition for \eref{ub_infty} to hold is when the law of $z$ has a point mass at $\ub$. The acquisition models described in \eref{logistic} and \eref{subset} in later sections are examples for which this condition is applicable.

The inequality in \ref{a:pos_corr} is also a natural requirement. To see this, we note that the data matrix $\mD_m$ in \eref{D_mtx} is the sample average of $m$ i.i.d. random rank-one matrices $\set{\mathcal{T}(y_i) \va_i \va_i^\top}_{i \le m}$. When the number of samples $m$ is large, this sample average should be ``close'' to the statistical expectation, \emph{i.e.},
\begin{equation}\label{eq:mtx_LLN}
\mD_m \approx \mathbb{E} (z_i \, \va_i \va_i^\top),
\end{equation}
where $z_i \bydef \mathcal{T}(y_i)$. To compute the above expectation, it will be convenient to assume that the underlying signal $\vxi = \nrm \ve_1$, where $\ve_1$ is the first vector of the canonical basis of $\R^n$. (This assumption can be made without loss of generality, due to the rotational invariance of the multivariate normal distribution.)  Correspondingly, we can partition each sensing vector into two parts, as
\begin{equation}\label{eq:a_splitting}
\va_i^\top = \begin{bmatrix} s_i & \vu_i^\top \end{bmatrix},
\end{equation}
so that $\va_i^\top \vxi = \nrm s_i$ and the conditional density of $y_i$ given $s_i$ is $f(y \,\vert\, \nrm s_i)$. Since $s_i, y_i$ and $z_i$ are all independent of $\vu_i$, 
\begin{align}
\mathbb{E} (z_i \, \va_i \va_i^\top) &= \mathbb{E}
\begin{bmatrix}
z_i s_i^2 & z_i s_i \vu_i^\top\\
z_i s_i \vu_i & z_i \vu_i \vu_i^\top
\end{bmatrix} \nonumber\\
&=
\begin{bmatrix}
\mathbb{E}\,z s^2 & \vzero\\
\vzero & \mathbb{E} z \,\mI_{n-1}
\end{bmatrix},\label{eq:expectation_mtx}
\end{align}
where $\mI_{n-1}$ is the identity matrix of size $(n-1)$. If the inequality $\mathbb{E}\,z s^2 > \mathbb{E}z$, as required in \ref{a:pos_corr}, indeed holds, the leading eigenvector of the expected matrix will be $\ve_1$, which is perfectly aligned with the target vector $\vxi$. Now since the data matrix $\mD_m$ is an approximation of the expectation, the sample eigenvector should also be an approximation of $\vxi$.

The above argument provides an intuitive but nonrigorous explanation for why the spectral initialization method would work. The approximation in \eref{mtx_LLN} can be made exact if the signal dimension $n$ is kept fixed and the number of measurement $m$ goes to infinity. However, we consider the case when $m$ and $n$ both tend to infinity, at a constant ratio $\alpha = m/n$ bounded away from $0$ and $\infty$. In this regime, the approximation in \eref{mtx_LLN} will not become an equality even if $m \rightarrow \infty$. As we will show, the correlation $\rho(\vxi_n, \vx_1^n)$ between the target vector $\vxi_n$ and the sample eigenvector $\vx_1^n$ will converge to a deterministic value $\rho(\alpha)$ that depends on the sampling ratio $\alpha$.

A notable exception to \eref{pos_corr} is when
\begin{equation}\label{eq:g_func}
g(s) \bydef \mathbb{E}_{z \vert s}(z \vert s) 
\end{equation}
is an odd function plus some arbitrary constant $C$. In this case, $\mathbb{E} zs^2 = \mathbb{E}[ g(s) s^2] = C$ and $\mathbb{E} z = \mathbb{E} g(s) = C$ and thus \eref{pos_corr} does not hold. In practice, this means that the spectral method will not be effective for acquisition models such as $z = \sign(s) + C$. We will revisit this point in \sref{conclusion} where we describe an alternative initialization scheme that can handle such cases.

A final remark before we present our main results: Since the eigenvector $\vx_1^n$ is always normalized, the spectral method cannot provide any information about the norm of $\vxi_n$. However, in many cases where the sensing vectors are drawn from certain random ensembles, there are simple methods to accurately estimate $\nrm = \norm{\vxi_n}$. We provide some discussions on how to do this in Appendix~\ref{appendix:norm}.

\subsection{Main Results: Asymptotic Characterizations}
\label{sec:main_results}

In this section, we summarize the main results of our work on an asymptotic characterization of the spectral method with Gaussian measurements. To state our results, we first need to introduce several helper functions. Let $s, z$ be the random variables defined in \eref{sy}. We consider two functions
\begin{equation}\label{eq:muf}
\muf(\lambda) \bydef \lambda \, \mathbb{E}\frac{ z s^2}{\lambda - z}
\end{equation}
and
\begin{equation}\label{eq:uf}
\uf_\alpha(\lambda) \bydef \lm \, \Big(1/\alpha + \mathbb{E} \frac{z}{\lambda-z}\Big),
\end{equation}
both defined on the open interval $(\ub, \infty)$, where $\ub$ is the bound in assumption~\ref{a:rvs}.
Within their domains, it is easy to check that both functions are convex. In particular, $\uf_\alpha(\lambda)$ achieves its minimum at a unique point denoted by 
\begin{equation}\label{eq:barlm}
\blm_\alpha \bydef \underset{\lambda > \tau}{\arg\,\min} \ \uf_\alpha(\lambda).
\end{equation}
Finally, let 
\begin{equation}\label{eq:lf}
\lf_\alpha(\lambda) \bydef \uf_\alpha\big(\lambda \vee \blm_\alpha\big)
\end{equation}
be a modification of $\uf_\alpha(\lambda)$. This new function is again defined for $\lambda \in (\ub, \infty)$.

\begin{theorem}\label{thm:cos2}
Under \ref{a:Gaussian} -- \ref{a:pos_corr}, the following hold:
\begin{enumerate}
\item There is a unique solution, denoted by $\lm^\ast_\alpha$, to the equation
\begin{equation}\label{eq:fixed_point}
\lf_\alpha(\lm) = \muf(\lm), \qquad \lm > \tau.
\end{equation}

\item As $n \rightarrow \infty$,
\begin{equation}\label{eq:cos2_general}
\rho(\vxi_n, \vx_1^n) \cip \begin{cases}
0, &\text{if } \uf'_\alpha(\lm^\ast_\alpha) < 0, \\
\frac{\uf'_\alpha(\lm^\ast_\alpha)}{\uf'_\alpha(\lm^\ast_\alpha) - \muf'(\lm^\ast_\alpha)}, &\text{if } \uf'_\alpha(\lm^\ast_\alpha) > 0,
\end{cases}
\end{equation}
where $\uf'_\alpha(\cdot)$ and $\muf'(\cdot)$ denote the derivatives of the two functions.

\item Let $\lambda_1^{\mD_m} \ge \lambda_2^{\mD_m}$ be the top two eigenvalues of $\mD_m$.
\begin{equation}\label{eq:l1l2}
\lambda_1^{\mD_m} \cip \lf_\alpha(\lm^\ast_\alpha) \quad \text{and} \quad \lambda_2^{\mD_m} \cip \lf_\alpha(\blm_\alpha)
\end{equation}
as $n \rightarrow \infty$. Moreover, $\lf_\alpha(\lm^\ast_\alpha) \ge \lf_\alpha(\blm_\alpha)$, with the inequality becoming strict \emph{if and only if} $\uf'_\alpha(\lm^\ast_\alpha) > 0$.

\end{enumerate}
\end{theorem}

\begin{remark}
The above theorem, whose proof is given in \sref{proof}, provides a complete asymptotic characterization of the performance of the spectral method. In particular, the theorem shows that the squared cosine similarity $\rho(\vxi_n, \vx_1^n)$ converges in probability to a deterministic value in the high-dimensional limit. Moreover, there exists a generic phase transition phenomenon: depending on the sign of the derivative $\uf'_\alpha(\cdot)$ at $\lm^\ast_\alpha$, the limiting value can be either zero (\emph{i.e.}, the uncorrelated phase) or strictly positive (\emph{i.e.}, the correlated phase). The computational complexity of the spectral method is also very different in the two phases. Within the uncorrelated phase, the gap between the top two leading eigenvalues, $\lm_1^{\mD_m}$ and $\lm_2^{\mD_m}$, diminishes to zero, making iterative algorithms such as power iterations increasingly difficult to converge. In contrast, within the correlated phase, the spectral gap converges to a positive value.
\end{remark}

\begin{remark}\label{rem:extension}
The results of this work were first reported in \cite{LuL:17a, LuL:17}. When this paper was under review, the results given in Theorem~\ref{thm:cos2} were further extended by Mondelli and Montanari in \cite{MondelliM:17}. In particular, these authors extended our asymptotic predictions from the real-valued case to the complex-valued case, and more importantly, they showed that the same predictions still hold under cases where the variable $z$ defined in assumption~\ref{a:rvs} can take negative values. See \cite[Lemma 2]{MondelliM:17} for details.
\end{remark}

It will be more convenient to characterize the phase transitions predicted by Theorem~\ref{thm:cos2} in terms of the sampling ratio $\alpha$. To do so, we first introduce a set $\Lambda$, containing all the zero-crossings of the function $\Delta(\lm) =  \lm \, \mathbb{E} \frac{z}{(\lm-z)^2} - \mathbb{E}\frac{zs^2}{\lm - z}$ on the open interval $(\ub, \infty)$. We can show that $\Lambda$ is always nonempty and that it contains a finite number of points (see Lemma~\ref{lemma:zeros} in \sref{zeros}). Let
\[
\lm_{c, \min} \bydef \underset{\lambda \in \Lambda}{\min} \, \lambda \quad \text{and} \quad \lm_{c, \max} \bydef \underset{\lambda \in \Lambda}{\max} \, \lambda
\]
denote the smallest and the largest elements in $\Lambda$, respectively. 

\begin{proposition}\label{prop:alpha_c}
Under \ref{a:Gaussian} -- \ref{a:pos_corr}, and as $n \rightarrow \infty$,
\[
\rho(\vxi_n, \vx_1^n) \cip \begin{cases}
0, &\text{if } \alpha < \alpha_{c,\min}, \\
\rho(\alpha), &\text{if } \alpha > \alpha_{c,\max},
\end{cases}
\] 
where
\begin{equation}\label{eq:alpha_min_max}
\alpha_{c,\min}^{-1} =  \mathbb{E} \frac{z^2}{(\lm_{c, \min} - z)^2}, \ \alpha_{c,\max}^{-1} = \mathbb{E} \frac{z^2}{(\lm_{c, \max} - z)^2},
\end{equation}
and $\rho(\alpha)$ is a function with the following parametric representation in terms of a parameter $\lambda$:
\begin{align}
1/\alpha &= \mathbb{E}\frac{zs^2 - z}{\lm - z} \label{eq:p_alpha}\\
1/\rho &= {1 + \Big(\mathbb{E}\frac{zs^2 - z}{\lm - z} - \mathbb{E}\frac{z^2}{(\lm - z)^2}\Big)^{-1}\mathbb{E}\frac{z^2 s^2}{(\lm - z)^2}},\label{eq:p_rho}
\end{align}
for all $\lambda > \lm_{c,\max}$. Moreover, $\rho(\alpha) \rightarrow 1$ as $\alpha \rightarrow \infty$.
\end{proposition}

\begin{remark}
In many of the signal acquisition models we have studied, the set $\Lambda$ contains exactly one element. In this case, $\lm_{c, \min} = \lm_{c, \max}$ and hence $\alpha_{c,\min} = \alpha_{c,\max}$. Consequently, the phase transition of the spectral method takes place at a single threshold value $\alpha_c$, which separates the uncorrelated phase from the correlated one. However, it is indeed possible to find cases for which $\alpha_{c,\min} < \alpha_{c,\max}$. This leads to a more complicated scenario, where a finite number of correlated and uncorrelated phases can alternatively take place within the interval $(\alpha_{c,\min}, \alpha_{c,\max})$. One such example is given in \sref{counter_example}.
\end{remark}

\subsection{Worked-Example: Binary Models}
\label{sec:one_bit}

To illustrate the results presented above, we consider here a special case where $z_i$ takes only binary values $\set{0,1}$. This situation naturally appears in problems such as logistic regression and one-bit quantized sensing, where the measurements $y_i \in \set{0, 1}$ and we can set $z_i = y_i$. For cases where the measurements $\set{y_i}$ are not necessarily binary, this type of one-bit model is still relevant whenever the preprocessing function $z = \mathcal{T}(x)$ generates binary outputs. The simplicity of this setting allows us to obtain closed-form expressions for the various quantities in Theorem~\ref{thm:cos2} and Proposition~\ref{prop:alpha_c}. 

To proceed, we first explicitly compute the functions $\muf(\lm)$ and $\uf_\alpha(\lm)$ defined in \sref{main_results} as
\[
\muf(\lm) = \frac{c \lm}{\lm - 1} \quad\text{and}\quad \uf_\alpha(\lm) = \lm\Big(1/\alpha + \frac{d}{\lm - 1}\Big),
\]
where 
\begin{equation}\label{eq:cd}
c \bydef \mathbb{E}\,{z s^2} \quad \text{and} \quad d \bydef \mathbb{E}\,{z}
\end{equation}
and both functions are defined on the interval $\lambda > 1$. The minimum of $\uf_\alpha(\lm)$ is achieved as $\blm_\alpha = 1 + \sqrt{\alpha d}$, and thus
\[
\lf_\alpha(\lm) = \begin{cases}
\lambda/\alpha + \lm d/(\lm-1), &\text{for } \lambda \ge 1 + \sqrt{\alpha d}\\
(\sqrt{d} + 1/\sqrt{\alpha})^2, &\text{for } 1 < \lambda < 1 + \sqrt{\alpha d}.
\end{cases}
\]
Solving equation \eref{fixed_point} and using \eref{cos2_general}, we get
\begin{equation}\label{eq:cos2_closed_form}
\rho(\vxi_n, \vx_1^n) \cip \begin{cases}
0, &\text{for } \alpha < \alpha_c, \\
\frac{\alpha - d/(c-d)^2}{\alpha + 1/(c-d)}, &\text{for } \alpha > \alpha_c,
\end{cases}
\end{equation}
where $\alpha_c = \frac{d}{(c-d)^2}$.  (Note that this result can also be obtained by invoking the parametric characterization of $\rho(\alpha)$ given in Proposition~\ref{prop:alpha_c}.) Finally, the asymptotic predictions \eref{l1l2} for the top two eigenvalues can be computed as
\begin{equation}\label{eq:lambda1_closed_form}
\lambda_1^{\mD_m} \cip \begin{cases}
(\sqrt{d} + 1/\sqrt{\alpha})^2, &\text{for } \alpha < \alpha_c, \\
c + \frac{c}{\alpha(c-d)}, &\text{for } \alpha > \alpha_c,
\end{cases}
\end{equation}
and
\begin{equation}\label{eq:lambda2_closed_form}
\lambda_2^{\mD_m} \cip (\sqrt{d} + 1/\sqrt{\alpha})^2
\end{equation}
for all $\alpha$. 

\begin{remark}
It is interesting to note that the asymptotic characterizations given in \eref{cos2_closed_form}, \eref{lambda1_closed_form} and \eref{lambda2_closed_form} are \emph{universal}, in the sense that they only depend on the two constants $c$ and $d$ defined in \eref{cd} but not on the exact details of the joint probability distributions of $s, y$ and $z$. Thus, for one-bit models, it suffices to compute the constants in \eref{cd}, which then completely determine the asymptotic performance of the spectral method.
\end{remark}

\subsection{Numerical Simulations}
\label{sec:numerical}

\begin{example}[Logistic regression]\label{ex:logistic}
Consider the case where $\set{y_i}$ are binary random variables generated according to the following conditional distribution:
\begin{equation}\label{eq:logistic}
f(y \,\vert\, \va^\top \vxi_n) \sim \text{Bernoulli}\left(\frac{1}{1 + \exp\set{-\va^\top \vxi_n + \beta}}\right),
\end{equation}
where $\beta$ is some constant. Let $z_i = \mathcal{T}(y_i) = y_i$. Since $z_i \in \set{0, 1}$, we just need to compute the constants $c$ and $d$ in \eref{cd}, after which we can use the closed-form expressions \eref{cos2_closed_form}, \eref{lambda1_closed_form} and \eref{lambda2_closed_form} to obtain the asymptotic predictions. In \fref{logistic:1} we compare the analytical prediction \eref{cos2_closed_form} of the squared cosine similarity with results of numerical simulations. In our experiment, we set the signal dimension to $n = 4096$. The norm of $\vxi_n$ is $\nrm = 3$, and $\beta = 6$. The sample averages and error bars (corresponding to one standard deviation) shown in the figure are calculated over 16 independent trials. We can see that the analytical predictions match numerical results very well. \fref{logistic:2} shows the top two eigenvalues. When $\alpha < \alpha_c$, the two eigenvalues are asymptotically equal, but they start to diverge as $\alpha$ becomes larger than $\alpha_c$. To clearly illustrate this phenomenon, we plot in the insert the eigengap $\lambda_1 - \lambda_2$ as a function of $\alpha$.
\end{example}

\begin{example}[Phase retrieval]\label{ex:phase}
In the second example, we consider the problem of phase retrieval, where
\[
y_i = (\va_i^\top \vxi_n)^2 + \sigma \omega_i.
\]
Here, $\omega_i \sim_{\text{i.i.d.}} \mathcal{N}(0, 1)$ and $\sigma \ge 0$ is the standard deviation of the noise. In \cite{Chen:2015eu}, the authors show that it is important to omit large values of $\set{y_i}$, and they propose to use the scheme in \eref{trimming} when constructing the data matrix $\mD_m$. A different strategy can be found in \cite{WangGY:2016}, where the authors propose to use
\begin{equation}\label{eq:subset}
z_i = \charfn_{\abs{y_i} > t^2}.
\end{equation}
In what follows, we shall refer to \eref{trimming} and \eref{subset} as the trimming algorithm and the subset algorithm, respectively. \fref{phase:1} shows the asymptotic performance of these two algorithms and compare them with numerical results ($n = 4096$ and 16 independent trials). The performance of the subset algorithm (for which we choose the parameter $t = 1.5$) can be characterized by the closed-form formula \eref{cos2_closed_form}. The trimming algorithm (for which we use $t = 3$) is more complicated as $z_i$ is no longer binary. We use the parametric characterization in Proposition~\ref{prop:alpha_c} to obtain its asymptotic performance. Again, our analytical predictions match numerical results. The performance of both algorithms clearly depends on the choice of the thresholding parameter $t$. To show this, we plot in \fref{phase:2} the critical phase transition points $\alpha_c$ of both algorithms as functions of $t$, at two different noise levels: $\sigma = 0$ and $\sigma = 2$. This points to the possibility of using our analytical prediction to optimally tune the algorithmic parameters and, more generally, to optimize the functional form of the preprocessing function $\mathcal{T}(\cdot)$. Indeed, the optimal design of $\mathcal{T}(\cdot)$ was obtained in a recent work \cite{LuoAL:19}, which leverages the asymptotic characterizations given here. Interestingly, under a mild technical condition, it is shown that there exists a simple fixed design that is \emph{uniformly optimal} over all sampling ratios; see \cite[Theorem~1]{LuoAL:19}.
%See \cite{MondelliM:17} for a design of $\mathcal{T}(\cdot)$ that is shown to achieve the optimal phase transition thresholds.
\end{example}

\begin{figure}
	\centering
	\subfigure[Squared cosine similarity]{\label{fig:logistic:1}
	\includegraphics[width=0.46\linewidth]{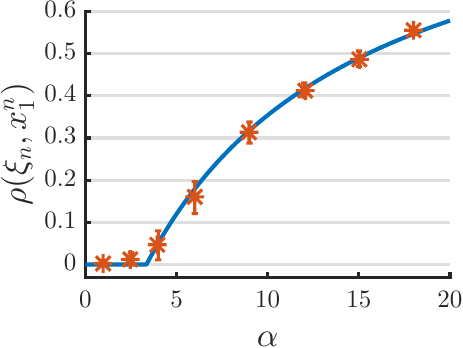}}
	\hspace{1ex}
	\subfigure[Top two eigenvalues]{\label{fig:logistic:2}
	\includegraphics[width=0.46\linewidth]{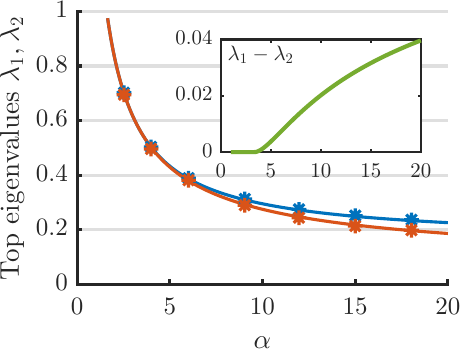}}
	\caption{Analytical predictions v.s. numerical simulations for the binary logistic model in \eref{logistic}. Numerical results are averaged over 16 independent trials.}
\end{figure}

\begin{figure}
	\centering
	\subfigure[Squared cosine similarity]{\label{fig:phase:1}
	\includegraphics[width=0.46\linewidth]{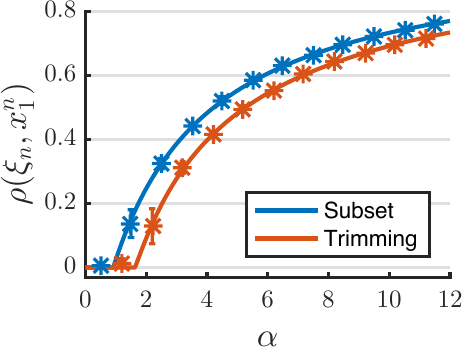}}
	\hspace{1ex}
	\subfigure[Phase transition points]{\label{fig:phase:2}
	\includegraphics[width=0.46\linewidth]{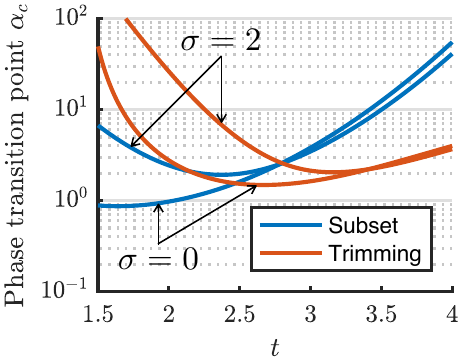}}
	\caption{(a) Analytical predictions v.s. numerical simulations for two different algorithms for phase retrieval in the noiseless setting (\emph{i.e.}, $\sigma = 0$). (b) The critical sampling ratio $\alpha_c$ of the two algorithms as functions of the threshold value $t$, at two different noise levels: $\sigma = 0$ and $\sigma = 2$.}
\end{figure}

\section{Proof of the Main Results}
\label{sec:proof}

In this section, we prove Theorem~\ref{thm:cos2}, which provides an exact characterization of the asymptotic performance of the spectral method for signal estimation. 

\subsection{Overview}

We first rewrite the data matrix $\mD_m$ in \eref{D_mtx} as
\begin{equation}\label{eq:D_cov}
\mD_m = \tfrac{1}{m}\mA \mZ \mA^\top,
\end{equation}
where $\mA = [\va_1, \va_2, \ldots, \va_m]$ is an $n \times m$ matrix of i.i.d. normal random variables and
\begin{equation}\label{eq:Z_mtx}
\mZ \bydef \diag\set{z_1, z_2, \ldots, z_m}
\end{equation}
is a diagonal matrix with entries $z_i = \mathcal{T}(y_i)$. Our goal boils down to studying the largest eigenvalue of $\mD_m$ and the associated eigenvector $\vx_1^n$. To simplify notation, we shall first assume that $\vxi_n = \nrm \ve_1$, with $\ve_1$ being the first vector in the canonical basis. 

%This assumption is not restrictive, since the multivariate normal distributions of the sensing vectors $\set{a_i}$ are rotationally invariant.

\begin{remark}
The non-null eigenvalues of $\mD_m$ are equal to those of a companion matrix
\[
\widetilde{\mD}_m = \tfrac{1}{m} \mZ^{1/2} \mA^\top \mA \mZ^{1/2},
\]
which bears strong resemblance to a sample covariance matrix. Limiting spectral distributions (LSDs) of sample covariance matrices have been extensively studied in random matrix theory; see for instance \cite{BaiS:2010} and the references given there. As a special case, when $\mZ$ is the identity matrix, the LSD of $\widetilde{\mD}_m$ is given by the classical Mar\v{c}enko-Pastur law \cite{Marcenko:1967}. Results for more general diagonal matrices $\mZ$ are also available \cite{SilversteinBai:1995}. However, in these studies, $\mZ$ and $\mA$ need to be independent. A challenge in our problem is that $\mZ$ and $\mA$ are \emph{correlated}. To see this, we partition each sensing vector $\va_i$ into two parts as in \eref{a_splitting}. We can then write
\begin{equation}\label{eq:A_mtx}
\mA = \begin{bmatrix}
\vs^\top\\
\mU
\end{bmatrix},
\end{equation}
where $\vs \bydef [s_1, s_2, \ldots, s_m]^\top$ is an $m$-dimensional Gaussian random vector, and $\mU$ is an $(n-1) \times m$ matrix consisting of i.i.d. standard normal random variables. Since $\vxi_n = \nrm \ve_1$, the diagonal elements of $\mZ$ are independent of $\mU$ but they do depend on $\vs$ through $y_i \sim f(y \, \vert \, \nrm s_i)$. Consequently, we cannot apply existing results on the LSD of sample covariance matrices to our case.
\end{remark}

Our proof of Theorem~\ref{thm:cos2} consists of two main ingredients. First, we will show in Proposition~\ref{prop:fixed_point} that $\lambda_1^{\mD_m}$ and $\rho(\vxi_n, \vx_1^n)$ can be obtained from a fixed-point equation involving a function $L_m(\mu)$, to be defined in \eref{L_m}, where $\mu > 0$ is an auxiliary variable. The main benefit of introducing the variable $\mu$ and the function $L_m(\mu)$ is that, for each $\mu > 0$, the above-mentioned correlation between $\mA$ and $\mZ$ can be effectively decoupled. This then allows us to obtain the second ingredient of our proof: using results from random matrix theory \cite{Benaych-Georges:2011, BaiY:12}, we show in \sref{L_asymptotic} that $L_m(\mu)$, under the assumption of Gaussian sensing vectors, will converge almost surely to a deterministic limit function as the dimension $n \rightarrow \infty$ (see Proposition~\ref{prop:spike_pop}).

% First,  This fixed-point characterization does not require the sensing vectors to be drawn from the Gaussian distribution. In fact, it is valid for any finite dimension $n$ and for any \emph{deterministic} sensing vectors.  There it will be clear why the introduction of $\mM(\mu)$ and $L(\mu)$ helps to decouple the above-mentioned correlation between $\mA$ and $\mZ$.

\subsection{A Fixed-Point Characterization}
\label{sec:deterministic}

By substituting \eref{A_mtx} into \eref{D_cov}, we can write $\mD_m$ in a more compact block-partitioned form as
\begin{equation}\label{eq:D_block}
\mD_m = \pmatnocross
\begin{pmat}[{|}]
a_m 	& \vq_m^\top \cr\-
\vq_m	& \mP_m \cr
\end{pmat},
\end{equation}
where
\begin{equation}\label{eq:a}
a_m \bydef \tfrac{1}{m}\sum_{i = 1}^m z_i s_i^2
\end{equation}
is a scalar that converges to $\mathbb{E}zs^2$ as $m \rightarrow \infty$,
\begin{equation}\label{eq:P}
\mP_m \bydef \tfrac{1}{m}\mU \mZ \mU^\top
\end{equation}
is a symmetric matrix, and
\begin{equation}\label{eq:q}
\vq_m \bydef \tfrac{1}{m}\mU \vv \ \text{ with }\ \vv \bydef [z_1 s_1, z_2 s_2, \ldots, z_m s_m]^\top.
\end{equation}
 
Next, we consider a parametric family of matrices $\set{\mP_m + \mu \hspace{0.5pt} \vq_m \vq_m^\top: {\mu > 0}}$, and let $L_m(\mu)$ denote their largest eigenvalues, \emph{i.e.},
\begin{equation}\label{eq:L_m}
L_m(\mu) \bydef \lm_1(\mP_m + \mu \hspace{0.5pt} \vq_m \vq_m^\top).
\end{equation}
In what follows, we show how to compute $\lambda_1^{\mD_m}$ and $\rho(\vxi_n, \vx_1^n)$ via a fixed-point equation involving $L_m(\mu)$. Since we assume that $\vxi = \nrm \, \ve_1$ and that the leading eigenvector $\vx_1^n$ is normalized, the quantity $\rho(\vxi_n, \vx_1^n)$ is equal to $(\ve_1^\top \vx_1^n)^2$, the squared magnitude of the first element of the eigenvector.

Our discussions below are general and they apply to any block-partitioned matrix in the form
\begin{equation}\label{eq:D_block_general}
\mD = \pmatnocross
\begin{pmat}[{|}]
a 	& \vq^\top \cr\-
\vq	& \mP \cr
\end{pmat}.
\end{equation}
Its components $a \in \R$, $\mP \in \R^{(n-1)\times(n-1)}$ and $\vq \in \R^{n-1}$ can be arbitrarily chosen, not necessarily defined as in \eref{a}, \eref{P} and \eref{q}. Our only requirements are that $\mP$ is a symmetric matrix and that $\norm{\vq} \neq 0$.

Let $\lambda_1^{\mP} \ge \lambda_2^{\mP} \ge \ldots \lambda_{n-1}^{\mP}$ be the set of eigenvalues of $\mP$, and let $\vw_1, \vw_2, \ldots, \vw_{n-1}$ be a corresponding set of orthonormal eigenvectors. Consider a function
\begin{equation}\label{eq:R_func}
R(\lambda) \bydef \vq^\top (\mP - \lambda \mI)^{-1} \vq = \sum_{i=1}^{n-1} \frac{(\vw_i^\top \vq)^2}{\lambda_i^{\mP} - \lambda},
\end{equation}
which has poles on those eigenvalues for which $\vw_i^\top \vq \neq 0$. In what follows, we restrict the domain of $R(\lambda)$ to 
\[
\lambda \in  \Big(\max\set{\lambda_i^{\mP}: \vw_i^\top \vq \neq 0}, \infty\Big).
\]
Within this open interval, $R(\lambda)$ is a well-defined smooth function. It increases monotonically from $-\infty$ to $0$, and thus it admits a functional inverse, denoted by $R^{-1}(x)$, for all $x < 0$. Similar to \eref{L_m}, we define
\[
L(\mu) = \lm_1(\mP + \mu \hspace{0.5pt} \vq \vq^\top)
\]
for all $\mu > 0$.

\begin{lemma}\label{lemma:L_rank_one}
Let $\mP$ be a symmetric matrix and $\vq$ a nonzero vector. Then, for each $\mu > 0$,
\begin{equation}\label{eq:L_rank_one}
L(\mu) = R^{-1}(-1/\mu) \vee \lambda_1^{\mP}.
\end{equation}
Moreover, $L(\mu)$ is a nondecreasing convex function with $\lim_{\mu \rightarrow \infty}L(\mu) = \infty$. It is differentiable everywhere on $(0, \infty)$ except at (up to) one point.
\end{lemma}
\begin{IEEEproof}
Since $\mP$ is diagonalizable by an orthonormal matrix, we can assume without loss of generality that $\mP$ is a diagonal matrix. In this case, we can simply write $R(\lambda) = \sum_i \frac{q_i^2}{\lambda_i^{\mP} - \lambda}$, and this function is defined on the open interval $(\max\set{\lambda_i^{\mP}: q_i \neq 0}, \infty)$. 

 Using the matrix determinant lemma \cite{Harville:2000}, we can compute the characteristic polynomial of $\mP + \mu \hspace{0.5pt} \vq \vq^\top$ as
\begin{align}
c(\lambda) &= \det(\lambda \mI - \mP - \mu \hspace{0.5pt} \vq \vq^\top) \nonumber\\
&= \det(\lambda \mI - \mP) - \mu \vq^\top \adj(\lambda \mI - \mP) \vq \label{eq:adj}\\
&= \prod_i (\lambda - \lambda_i^{\mP}) - \mu \sum_i q_i^2 \prod_{j \neq i} (\lambda - \lambda_j^{\mP}).\label{eq:c_poly}
\end{align}
In \eref{adj}, $\adj(\cdot)$ stands for the adjugate of a matrix. To reach \eref{c_poly}, we have used the fact that, for any diagonal matrix $\mA = \diag\set{d_1, d_2, \ldots, d_{n-1}}$, $\adj(\mA) = \diag\set{\prod_{j \neq 1} d_j, \prod_{j \neq 2} d_j, \ldots, \prod_{j \neq n-1} d_j}$.

Partition the set $\set{1, 2, \ldots, n-1}$ into two subsets:
\begin{equation}\label{eq:subsets}
\mathcal{I}_1 = \set{i: q_i \neq 0} \quad \text{and} \quad \mathcal{I}_2 = \set{i: q_i = 0}.
\end{equation}
We observe that the characteristic polynomial can be factored into $c(\lambda) = c_1(\lambda) c_2(\lambda)$, where $c_2(\lambda) = \prod_{i \in \mathcal{I}_2} (\lambda - \lambda_i^{\mP})$ and
\[
c_1(\lambda) = \prod_{i \in \mathcal{I}_1} (\lambda - \lambda_i^{\mP}) - \mu \sum_{i \in \mathcal{I}_1} q_i^2 \prod_{j \in \mathcal{I}_1 \setminus  i} (\lambda - \lambda_j^{\mP}).
\]
It is possible that the second subset $\mathcal{I}_2$ is empty, in which case $c_2(\lambda)$ is understood to be equal to $1$, but $\mathcal{I}_1$ is never empty, since $\vq \neq \vzero$. Next, we study the largest root of the polynomial $c_1(\lambda)$. For any $\lambda > \max\set{\lambda_i^{\mP}: i \in \mathcal{I}_1}$, we can write
\begin{align}
c_1(\lambda) &=  \Big(1 - \mu \sum_{i \in \mathcal{I}_1} \frac{q_i^2}{\lambda - \lambda_i^{\mP}}\Big) \prod_{i \in \mathcal{I}_1} (\lambda - \lambda_i^{\mP})\nonumber\\
&= \big(1 + \mu R(\lambda)\big) \prod_{i \in \mathcal{I}_1} (\lambda - \lambda_i^{\mP}).\label{eq:c2lm}
\end{align}
Recall that $R(\lm)$ is the function defined in \eref{R_func} and $R^{-1}(\cdot)$ is its functional inverse. It follows from \eref{c2lm} that $R^{-1}(-1/\mu)$ is the only root of $c_1(\lambda)$ in the interval $(\max\set{\lambda_i^{\mP}: i \in \mathcal{I}_1}, \infty)$, and therefore it is also the largest root. Due to the factorization $c(\lambda) = c_1(\lambda) c_2(\lambda)$, we have
\[
L(\mu) = R^{-1}(-1/\mu) \vee \max\set{\lambda_i: i \in \mathcal{I}_2}.
\]
Finally, since $R^{-1}(-1/\mu) > \max\set{\lambda_i: i \in \mathcal{I}_1}$, we reach the formula in \eref{L_rank_one}.

By construction, $R^{-1}(-1/\mu)$ is strictly increasing and $\lim_{\mu \rightarrow \infty} R^{-1}(-1/\mu) = \infty$. It is also differentiable everywhere on $(0, \infty)$. It follows that $L(\mu)$ is nondecreasing with $L(\infty) = \infty$, and that the function is differentiable everywhere except for at most one point $\mu_0$, which, if it exists, must satisfy the identity $R^{-1}(-1/\mu_0) = \lambda_1^{\mP}$. Finally, the convexity of $L(\mu)$ follows from the fact that it is the maximum of a set of linear functions, as $L(\mu) = \lambda_1(\mP + \mu \hspace{0.5pt} \vq \vq^\top) = \max_{\vx: \, \norm{\vx} = 1} \vx^\top (\mP + \mu \hspace{0.5pt} \vq \vq^\top) \vx$.
\end{IEEEproof}

Given a block-partitioned matrix $\mD$, the following proposition shows that its leading eigenvalue $\lambda_1^{\mD}$ and the squared cosine similarity $(\ve_1^\top \vx_1)^2$ can be obtained from the function $L(\mu)$.

\begin{proposition}\label{prop:fixed_point}
Let $\mu^\ast > 0$ be the unique solution to the fixed-point equation
\begin{equation}\label{eq:mu_fixed_point}
\mu = (L(\mu) - a)^{-1}.
\end{equation}
 Then, $\lambda_1^{\mD} = L(\mu^\ast)$ and
\begin{equation}\label{eq:interval1}
(\ve_1^\top \vx_1)^2 \in \left[\frac{\partial_{-}L(\mu^\ast)}{\partial_{-}L(\mu^\ast) + (1 / \mu^\ast)^2}, \frac{\partial_{+}L(\mu^\ast)}{\partial_{+}L(\mu^\ast) + (1 / \mu^\ast)^2}\right],
\end{equation}
where $\partial_{-}L(\mu)$ and $\partial_{+}L(\mu)$ denote the left and right derivatives of $L(\mu)$, respectively. In particular, if $L(\mu)$ is differentiable at $\mu^\ast$, then
\begin{equation}\label{eq:characterization}
(\ve_1^\top \vx_1)^2 = \frac{L'(\mu^\ast)}{L'(\mu^\ast) + (1 / \mu^\ast)^2}.
\end{equation}
\end{proposition}
\begin{remark}\label{rem:mu}
We prove this result in Appendix~\ref{appendix:fixed_point}. Note that \eref{mu_fixed_point} is equivalent to
\begin{equation}\label{eq:mu_fixed_point2}
L(\mu) = a + 1 / \mu.
\end{equation}
Since $L(\mu)$ is nondecreasing with $L(\infty) = \infty$ whereas $a + 1/\mu$ decreases monotonically from $\infty$ to $0$, the equation \eref{mu_fixed_point2}, and thus \eref{mu_fixed_point}, always admits one and only one solution. Moreover, by Lemma~\ref{lemma:L_rank_one}, $L(\mu)$ is a convex function, and therefore its left and right derivatives always exist.
\end{remark}

\subsection{Asymptotic Limit of $L_m(\mu)$}
\label{sec:L_asymptotic}

The characterization given in Proposition~\ref{prop:fixed_point} is valid for any block-partitioned matrix in the form of \eref{D_block_general}. When applied to the specific case of our data matrix in \eref{D_block}, with its components $a_m$, $\mP_m$ and $\vq_m$  defined as in \eref{a}, \eref{P} and \eref{q}, this result provides a very general \emph{deterministic} characterization of the performance of the spectral method that is valid for any finite dimension $n$ and for any sensing vectors.

Next, we specialize to the case of i.i.d. Gaussian sensing vectors and show that $L_m(\mu)$ converges almost surely to a deterministic function as $m,n \rightarrow \infty$. To that end, we note that $L_m(\mu)$ is the leading eigenvalue of
\begin{equation}\label{eq:umu}
\mP_m + \mu \hspace{0.5pt} \vq_m \vq_m^\top = \tfrac{1}{m} \mU \mM_m \mU^\top,
\end{equation}
where
\begin{equation}\label{eq:M_mtx}
\mM_m \bydef \mZ + \tfrac{\mu}{m} \vv \vv^\top
\end{equation}
is a rank-one perturbation of the diagonal matrix $\mZ$ given in \eref{Z_mtx}. Since $\mU$ and $\mM_m$ are independent, we first study the spectrum of $\mM_m$.

Let $\lambda_1^{\mM_m} \ge \lambda_2^{\mM_m} \ge \ldots \ge \lambda_m^{\mM_m}$ be the set of eigenvalues of $\mM_m$ in descending order. Let
\[
f^{\mM_m}(\lambda)  \bydef \frac{1}{m-1} \sum_{i=2}^{m} \delta(\lambda - \lambda_i^{\mM_m})
\]
be the empirical spectral measure of the last $m-1$ eigenvalues.

\begin{proposition}\label{prop:bulk_spike}
Fix $\mu > 0$. As $m, n \rightarrow \infty$, the empirical spectral measure $f^{\mM_m}(\lambda)$ converges almost surely to the probability law of the random variable $z$. Meanwhile,
\begin{equation}\label{eq:lmu_limit}
\lambda_1^{\mM_m} \asc Q^{-1}(1/\mu),
\end{equation}
where $Q^{-1}(\cdot)$ is the functional inverse of the function
\begin{equation}\label{eq:q_lm}
Q(\lm) = \mathbb{E}\frac{z^2 s^2}{\lambda-z}.
\end{equation}
The domain of $Q(\lm)$ is the open interval $(\tau, \infty)$, with $\tau$ being the upper bound of the support of the probability law of $z$.
\end{proposition}

\begin{remark}
By construction, $Q(\lm)$ is a continuous and strictly decreasing function with $Q(\infty) = 0$. Assumption~\ref{a:infty} further guarantees that $\lim_{\lm \rightarrow \ub^{+}} Q(\lm) = \infty$. Thus, $Q(\lm)$ admits a functional inverse and that $Q^{-1}(1/\mu)$ is well-defined for all $\mu > 0$. 
\end{remark}

According to assumption~\ref{a:rvs} stated in \sref{assumptions}, the law of $z$ is supported within the interval $[0, \tau]$. The above proposition, whose proof can be found in Appendix~\ref{appendix:bulk_spike}, shows that the spectrum of $\mM_m$ consists of two parts: a ``bulk spectrum'' of $m-1$ eigenvalues supported within $[0, \tau]$ and a single spiked eigenvalue $\lambda_1^{\mM_m}$ well separated from the bulk. This setting is a generalization of the classical spiked population model \cite{Johnstone:2001}. Adapting the results given in \cite{BaiY:12} (see also \cite{Benaych-Georges:2011} for related results under more general settings), we thus reach the second important ingredient of our proof of Theorem~\ref{thm:cos2}, characterizing the asymptotic limit of $L_m(\mu)$.

\begin{proposition}\label{prop:spike_pop}
For each fixed $\mu > 0$,
\begin{equation}\label{eq:Lmu}
L_m(\mu) \asc \lf_\alpha(Q^{-1}(1/\mu)),
\end{equation}
where $\lf_\alpha(\cdot)$ is the function defined in \eref{lf} and $Q^{-1}(1/\mu)$ is the limit value in \eref{lmu_limit}.
\end{proposition}
\begin{IEEEproof}
Recall from \eref{umu} that $L_m(\mu)$ is the leading eigenvalue of $\tfrac{1}{m} \mU \mM_m \mU^\top$. Since $\mU$ and $\mM_m$ are independent, and since $\mU$ is a Gaussian random matrix with a rotationally invariant distribution, we can equivalently study the leading eigenvalue of the following matrix
\begin{equation}\label{eq:ulmu}
\tfrac{1}{m} \mU \diag\set{\lm_1^{\mM_m}, \lm_2^{\mM_m}, \ldots, \lm_m^{\mM_m}} \mU^\top.
\end{equation}
Proposition~\ref{prop:bulk_spike} shows that $\set{\lm_i^{\mM}: i \ge 2}$ form a bulk spectrum, which converges to the law of $z$ as $m \rightarrow \infty$, whereas $\lm_1^{\mM}$ converges to a ``spike'' $\lm_\mu = Q^{-1}(1/\mu)> \ub$, which is separated from the bulk.

The asymptotic limits of extreme sample eigenvalues of matrices in the form of \eref{ulmu} have been studied in \cite{Benaych-Georges:2011, BaiY:12}. In our proof, we use the asymptotic characterization given in \cite{BaiY:12}. Key to this asymptotic analysis is the function $\uf_\alpha(\lm)$ defined\footnote{We have adapted the original definition of $\uf_\alpha(\lm)$ in \cite[eq. (3.2)]{BaiY:12} because our matrix in \eref{ulmu} has a slightly different scaling from the one considered in \cite{BaiY:12}.} in \eref{uf}. The asymptotic behaviors of the leading sample eigenvalue turn out to depend on the sign of $\uf'_\alpha(\lm)$ at the point $\lm_{\mu}$:

In particular, applying \cite[Theorem 4.1]{BaiY:12}, we have 
\begin{equation}\label{eq:distant}
L_m(\mu) \asc \uf_\alpha(\lm_\mu) \quad \text{if }\uf'_\alpha(\lm_\mu) > 0.
\end{equation}
 The case when $\uf'_\alpha(\lm_\mu) \le 0$ is covered in \cite[Theorem 4.2]{BaiY:12}. Adapting that result to our specific setting, we have
\begin{equation}\label{eq:close}
L_m(\mu) \asc \min_{\lm > \ub} \uf_\alpha(\lm) \quad \text{if }\uf'_\alpha(\lm_\mu) \le 0.
\end{equation}
As an equivalent form, we can write $\uf_\alpha(\lm) = \mathbb{E} z + \frac{\lm}{\alpha} + \mathbb{E}\frac{z^2}{\lm - z}$. From this, we can easily check that $\uf_\alpha(\lm)$ is a convex function and that it admits a unique minimum within its domain $(\ub, \infty)$. It follows that the two separate cases in \eref{distant} and \eref{close} can be more compactly written as $L_m(\mu) \asc \lf_\alpha(\lm_\mu)$, where $\lf_\alpha(\cdot)$ is the modified function defined in \eref{lf}.
\end{IEEEproof}

\subsection{Proof of Theorem~\ref{thm:cos2}}

We are now ready to prove our asymptotic characterizations given in Theorem~\ref{thm:cos2}. Since the sensing vectors $\va_i$ are drawn from the rotationally invariant multivariate normal distribution, the quantity $\rho(\vxi_n, \vx_1^n)$ for a general vector $\vxi_n$ (with $\norm{\vxi_n} = \nrm$) and $\rho(\nrm \ve_1, \vx_1^n)$ for the special case $\vxi_n = \nrm \ve_1$ have exactly the same probability distribution. In what follows, we will carry out the proof by assuming that the target vector $\vxi_n = \nrm \ve_1$. By showing that $\rho(\nrm \ve_1, \vx_1^n)$ converges to the right-hand side of \eref{cos2_general} \emph{almost surely}, the convergence to the same limit in probability for a general $\vxi_n$ then follows as an immediate consequence.

To start, we use the deterministic characterization given in Proposition~\ref{prop:fixed_point}. For each $m \ge 1$, let $\mu_m$ be the unique fixed-point of \eref{mu_fixed_point}. Equivalently, $\mu_m$ satisfies the identity
\[
L_m(\mu_m) - 1 / \mu_m = a_m.
\]
By Proposition~\ref{prop:spike_pop}, for every fixed $\mu$,
\begin{equation}\label{eq:L_limit_2}
L_m(\mu) - 1/\mu \asc \lf_\alpha(Q^{-1}(1/\mu)) - 1/\mu
\end{equation}
as $m \rightarrow \infty$. Since $L_m(\mu)$ and $\lf_\alpha(\mu)$ are nondecreasing, the two functions on both sides of \eref{L_limit_2} are strictly increasing. This condition, together with the fact that $a_m \asc \mathbb{E}{zs^2} $, allows us to apply Lemma~\ref{lemma:inverse} in Appendix~\ref{appendix:auxiliary} to conclude $\mu_m \asc \mu^\ast$, where $\mu^\ast$ is the unique point such that
\begin{equation}\label{eq:mu_characterization}
\lf_\alpha(Q^{-1}(1/\mu^\ast)) = \mathbb{E}{zs^2} + 1 / \mu^\ast.
\end{equation}

To determine the asymptotic behavior of the leading eigenvector $\vx_1^n$, we use the characterization given in \eref{interval1}. Since $\set{L_m(\mu)}$ are convex functions, we apply Lemma~\ref{lemma:derivative} in Appendix~\ref{appendix:auxiliary}. In particular, if $\lf_\alpha(Q^{-1}(1/\mu))$ is differentiable at $\mu = \mu^\ast$, that lemma gives us
\[
\partial_- L_m(\mu_m) \asc \eval{\dod{\lf_\alpha(Q^{-1}(1/\mu))}{\mu}}_{\mu^\ast} = \frac{-\lf'_\alpha(Q^{-1}(1/\mu^\ast))}{Q'(Q^{-1}(1/\mu^\ast))(\mu^\ast)^2}
\]
and similarly
\[
\partial_+ L_m(\mu_m) \asc \frac{-\lf'_\alpha(Q^{-1}(1/\mu^\ast))}{Q'(Q^{-1}(1/\mu^\ast))(\mu^\ast)^2}.
\]
Substituting these limits into \eref{interval1}, we get
\begin{equation}\label{eq:cos2_qinverse}
(\ve_1^\top \vx_1^n)^2 \asc \frac{\lf'_\alpha(Q^{-1}(1/\mu^\ast))}{\lf'_\alpha(Q^{-1}(1/\mu^\ast)) - Q'(Q^{-1}(1/\mu^\ast))}.
\end{equation}

To simplify the above expression, we introduce a change of variable, writing $\lm = Q^{-1}(1/\mu)$. In particular, $\lm^\ast = Q^{-1}(1/\mu^\ast)$. Using the characterization \eref{mu_characterization} and recalling the definition of $Q(\lm)$ in \eref{q_lm}, we get
\begin{equation}\label{eq:lm_characterization}
\lf_\alpha(\lm^\ast) = \mathbb{E} zs^2 + Q(\lm^\ast) = \muf(\lm^\ast),
\end{equation}
where $\muf(\cdot)$ is defined in \eref{muf}. By their constructions, it is easily checked that $\lf_\alpha(\lm)$ is a nondecreasing continuous function on $(\ub, \infty)$ whereas $\muf(\lm)$ is a strictly decreasing continuous function. Moreover, by assumption~\ref{a:infty}, $\lim_{\lm \rightarrow \ub^+} \muf(\lm) = \infty$. Thus, the existence of $\lm^\ast$ satisfying \eref{lm_characterization} and its uniqueness are guaranteed. Substituting $\lm^\ast = Q^{-1}(1/\mu^\ast)$ into \eref{cos2_qinverse} gives us
\[
(\ve_1^\top \vx_1^n)^2 \asc \frac{\lf'_\alpha(\lm^\ast)}{\lf'_\alpha(\lm^\ast) - \muf'(\lm^\ast)},
\]
where we have also used the fact that $Q'(\lm) = \muf'(\lm)$. To reach the characterization \eref{cos2_general} given in the theorem, we just need to note that, by its definition in \eref{lf}, $\lf'_\alpha(\lm) = \uf'_\alpha(\lm)$ if $\uf'_\alpha(\lm) > 0$ and $\lf'_\alpha(\lm) = 0$ if $\uf'_\alpha(\lm) < 0$.

Next, we characterize the first two eigenvalues $\lm_1^{\mD_m}$ and $\lm_2^{\mD_m}$. By Proposition~\ref{prop:fixed_point}, the leading eigenvalue $\lm_1^{\mD_m} = L_m(\mu_m)$. Since $\mu_m \asc \mu^\ast$, applying Lemma~\ref{lemma:inverse} stated in Appendix~\ref{appendix:auxiliary} leads to
\[
\lm_1^{\mD_m} \asc \lf_\alpha(Q^{-1}(1/\mu^\ast)) = \lf_\alpha(\lm^\ast).
\]
Recall from \eref{D_block} that $\mP_m$ is a principal submatrix of $\mD_m$ obtained by deleting the first row and column of $\mD_m$. It follows from the standard Cauchy interlacing theorem (see, \emph{e.g.}, \cite[Theorem 4.3.8]{HornJ:85}) that
\begin{equation}\label{eq:interlacing}
\lm_2^{\mP_m} \le\lm_2^{\mD_m} \le \lm_1^{\mP_m}
\end{equation}
Applying \cite[Lemma 3.1]{BaiY:12} (which is due to \cite{Silverstein:1995}), the upper edge of the support of the limiting spectral density of $\mP_m$ is given by
\[
\min_{\lm > \ub} \uf_\alpha(\lm) = \lf_\alpha(\overline{\lm}_\alpha),
\]
where $\overline{\lm}_\alpha$ is the minimizing point defined in \eref{barlm}. It follows that $\lm_2^{\mP_m} \asc \lf_\alpha(\overline{\lm}_\alpha)$ and $\lm_1^{\mP_m} \asc \lf_\alpha(\overline{\lm}_\alpha)$, and thus 
\[
\lm_2^{\mD_m} \asc \lf_\alpha(\overline{\lm}_\alpha)
\]
by the interlacing inequalities in \eref{interlacing}. Finally, by the constructions of $\uf_\alpha(\lm)$ and $\lf_\alpha(\lm)$, we have $\lf_\alpha(\lm) > \lf_\alpha(\overline{\lm}_\alpha)$ if and only if $\uf'_\alpha(\lm) > 0$, and the proof is complete.

%Substituting the limiting function $\lf_\alpha(\lm_\mu)$ into \eref{characterization}, and after some simple manipulations, we can (formally) reach the expression for the limiting squared cosine similarity given in \eref{cos2_general}. To rigorously prove the convergence, we still need to extend the pointwise convergence in \eref{Lmu} to uniform convergence over compact intervals and to establish the convergence of the derivatives. Details can be found in \cite{LuL:17}.

\section{Sampling Ratios and Phase Transitions}
\label{sec:phase}

In this section, we study the phase transition phenomena characterized in Theorem~\ref{thm:cos2} in more detail. In particular, we prove Proposition~\ref{prop:alpha_c} (as stated in \sref{main_results}), which specifies the phase transitions and the asymptotic limits of the cosine similarities in terms of the sampling ratio $\alpha$.

\subsection{Critical Sampling Ratios}
\label{sec:zeros}

By Theorem~\ref{thm:cos2}, whether the leading eigenvector $\vx_1^n$ is asymptotically correlated or uncorrelated with the target vector $\vxi_n$ depends on the sign of the derivative $\uf'_\alpha(\lm)$ evaluated at a point $\lm^\ast_\alpha$. And this point is uniquely defined through the equation $\lf_\alpha(\lm^\ast_\alpha) = \muf(\lm^\ast_\alpha)$. Let $\blm_\alpha$, defined in \eref{barlm}, be the point at which the strictly convex function $\uf_\alpha(\lm)$ achieves its minimum. Calculating the derivative of $\uf_\alpha(\lm)$ and setting it to zero, we get
\begin{equation}\label{eq:blm_alpha}
{1}/{\alpha} = \mathbb{E} \frac{z^2}{(\blm_\alpha - z)^2}.
\end{equation}

By the construction of the function $\lf_\alpha(\lm)$ in \eref{lf} and by the monotonicity of $\muf(\lm)$, we can conclude that $\uf'(\lm^\ast_\alpha) > 0$ \emph{if and only if} 
\[
\uf_\alpha(\blm_\alpha) < \muf(\blm_\alpha).
\]
Substituting \eref{blm_alpha} into \eref{uf} gives us $\uf_\alpha(\blm_\alpha) = \blm_\alpha^2 \, \mathbb{E} \frac{z}{(\blm_\alpha - z)^2}$. 
Thus, transitions between the correlated and uncorrelated phases take place exactly at the zero-crossings of the function
\begin{equation}\label{eq:delta_func}
\Delta(\lm) = \mathbb{E}\frac{\lm z}{(\lm-z)^2} - \mathbb{E}\frac{zs^2}{\lm -z},
\end{equation}
where $\Delta(\lm)$ is obtained by removing a common factor $\blm_\alpha$ from the difference $\uf_\alpha(\blm_\alpha) - \muf(\blm_\alpha)$ and by writing $\blm_\alpha$ simply as $\lm$. Let $\Lambda$ be the set consisting of all the zero-crossings of $\Delta(\lm)$ within the open interval $(\ub, \infty)$. Using \eref{blm_alpha}, we can then establish a one-to-one mapping between points in $\Lambda$ and a set of critical values of the sampling ratios.

\begin{lemma}\label{lemma:zeros}
The set $\Lambda$ is nonempty. It contains a finite number of points, denoted by $\lm_{c,1} \le \lm_{c,2} \le \ldots \le \lm_{c, r}$ for some $r \ge 1$. Moreover,
\begin{equation}\label{eq:zc_bound}
\lm_{c,r} \le \frac{\tau}{1-\sqrt{\mathbb{E} z / \mathbb{E} zs^2}}.
\end{equation}
\end{lemma}
\begin{IEEEproof}
We first show that $\Lambda$ is nonempty. For $\lm > \ub$, applying the Cauchy-Schwartz inequality gives us
\[
\begin{aligned}
\Delta(\lm) &\ge \mathbb{E}\frac{\lm z}{(\lm-z)^2} - \Big(\mathbb{E}\frac{z^2}{(\lm - z)^2}\Big)^{1/2} (\mathbb{E} \, s^4)^{1/2}\\
	&\ge \mathbb{E}\frac{\ub z}{(\lm-z)^2} - \Big(\mathbb{E}\frac{3 \ub z}{(\lm-z)^2}\Big)^{1/2}.
\end{aligned}
\]
By assumption~\ref{a:infty}, $\mathbb{E}\frac{z}{(\lm-z)^2} \rightarrow \infty$ as $\lm$ approaches $\ub$ from the right. Thus, we have
\begin{equation}\label{eq:g_left}
\lim_{\lm \rightarrow \ub^+} \Delta(\lm) = \infty.
\end{equation}

To study the function $\Delta(\lm)$ as $\lm \rightarrow \infty$, we note that
\begin{equation}\label{eq:g_right}
\Delta(\lm) \le \frac{1}{\lm}\left(\frac{\lm^2}{(\lm - \ub)^2} \mathbb{E} z - \mathbb{E} zs^2\right).
\end{equation}
By assumption~\ref{a:pos_corr}, $\mathbb{E} z < \mathbb{E} zs^2$. We can then conclude from inequality \eref{g_right} that
\begin{equation}
\Delta(\lm) < 0, \quad \text{for all sufficiently large } \lm.
\end{equation}
Since $\Delta(\lm)$ is a continuous function, \eref{g_left} and \eref{g_right} imply that there must exist at least one zero-crossing. 

Next, we show the upper bound given in \eref{zc_bound}. For any $\lm_c \in \Lambda$, we have from \eref{delta_func} that
\begin{equation}\label{eq:lm_c}
\lm_c = \Big(\mathbb{E}\frac{zs^2}{\lm-z}\Big) \Big(\mathbb{E}\frac{z}{(\lm-z)^2}\Big)^{-1}.
\end{equation}
By assumption~\ref{a:rvs}, $z$ is bounded within $[0, \ub]$. It follows that
\[
\mathbb{E}\frac{zs^2}{\lm-z} \ge \frac{\mathbb{E}zs^2}{\lm} \quad \text{and} \quad \mathbb{E}\frac{z}{(\lm-z)^2} \le \frac{\mathbb{E} z}{(\lm - \ub)^2}.
\]
Substituting the above inequalities into \eref{lm_c} gives us $(\mathbb{E} z) \lm_c^2 \ge (\mathbb{E}zs^2) (\lm_c - \ub)^2$, which, after some simple manipulations, leads to the upper bound given in \eref{zc_bound}.

Finally, to show that $\Lambda$ is a finite set, we  extend $\Delta(\lm)$ in \eref{delta_func} to the complex domain $\set{\lm \in \C: \operatorname{Re}(\lm) > \ub}$. Since $\Delta(\lm)$ is analytic and it is not zero everywhere, by the principle of permanence, it has at most a finite number of zeros in the bounded domain $(\ub, \frac{\tau}{1-\sqrt{\mathbb{E} z / \mathbb{E} zs^2}})$.
\end{IEEEproof}

\subsection{Proof of Proposition~\ref{prop:alpha_c}}

Write $\lm_{c,\min} = \lm_{c, 1}$ and $\lm_{c, \max} = \lm_{c, r}$. The corresponding critical sampling ratios $\alpha_{c,\min}$ and $\alpha_{c, \max}$, as defined in \eref{alpha_min_max},  are obtained through the one-to-one mapping given in \eref{blm_alpha}.

Fix $\alpha < \alpha_{c, \min}$. By the monotonicity of the mapping \eref{blm_alpha}, the corresponding $\blm_\alpha$ is strictly less than the smallest zero-crossing  point $\lm_{c, 1}$. From the proof of Lemma~\ref{lemma:zeros}, we conclude that $\Delta(\blm_\alpha) > 0$, and thus $\lf_\alpha(\cdot)$ and $\muf_\alpha(\cdot)$ intersects at a point $\lm^\ast_\alpha < \blm_\alpha$. This implies that $\uf'_\alpha(\lm^\ast_\alpha) < 0$ and thus Theorem~\ref{thm:cos2} gives us
\[
\rho(\vxi_n, \vx_1^n) \cip 0.
\]

Now fix $\alpha > \alpha_{c, \max}$, in which case $\blm_\alpha > \lm_{c, \max}$. Since $\Delta(\blm_\alpha) < 0$, we must have $\lm^\ast_\alpha > \blm_\alpha$ and thus $\uf'_\alpha(\lm^\ast_\alpha) > 0$. To derive the parametric form of $\rho(\alpha)$ given in the statement of the proposition, we note that
\[
\lf_\alpha(\lm) = \uf_\alpha(\lm) \quad \text{for all } \lm > \blm_\alpha.
\]
Thus, the equation $\lf_\alpha(\lm^\ast_\alpha) = \muf(\slma)$ becomes $\uf_\alpha(\lm^\ast_\alpha) = \muf(\slma)$. Using the explicit definitions of these functions given in \eref{muf} and \eref{uf}, we get
\begin{equation}\label{eq:p1}
1/\alpha = \mathbb{E}\frac{zs^2 - z}{\lm^\ast_\alpha - z}.
\end{equation}
We can also explicitly compute
\begin{align}
\uf'_\alpha(\lm^\ast_\alpha) &= 1/\alpha - \mathbb{E}\frac{z^2}{(\lm^\ast_\alpha -z)^2} \label{eq:p2_alpha}\\
	&= \mathbb{E}\frac{zs^2 - z}{\lm^\ast_\alpha - z} - \mathbb{E}\frac{z^2}{(\lm^\ast_\alpha -z)^2}.\label{eq:p2}
\end{align}
Similarly, we can write
\begin{equation}\label{eq:p3}
\muf'(\lm^\ast_\alpha) = - \mathbb{E}\frac{z^2 s^2}{(\lm^\ast_\alpha - z)^2}.
\end{equation}
Substituting \eref{p2} and \eref{p3} into the asymptotic characterization \eref{cos2_general} gives us \eref{p_rho}, which, together with \eref{p1}, provides a parametric representation of the function $\rho(\alpha)$.

Finally, we show that $\rho(\alpha) \rightarrow 1$ as $\alpha \rightarrow \infty$. From \eref{p1} and after some simple manipulations, we have
\[
{\lm^\ast_\alpha}/{\alpha} = \mathbb{E}(zs^2 - z) + \mathbb{E}\frac{z^2s^2 - z^2}{\lm^\ast_\alpha - z},
\]
Since $\lm^\ast_\alpha \rightarrow \infty$ as $\alpha \rightarrow \infty$, the above formula gives us
\[
\slma = \mathcal{O}(\alpha),
\]
where the leading coefficient $\mathbb{E}(zs^2 - z)$ is positive by assumption~\ref{a:pos_corr}. By the boundedness of $z$,
\[
\frac{\mathbb{E} z^2}{(\slma)^2} \le \mathbb{E}\frac{z^2}{(\slma - z)^2} \le \frac{\mathbb{E} z^2}{(\slma - \ub)^2},
\]
and thus $\mathbb{E}\frac{z^2}{(\slma - z)^2} = \mathcal{O}(1/\alpha^2)$. It follows from \eref{p2_alpha} that $\uf'_\alpha(\slma) = \mathcal{O}(1/\alpha)$. Similarly, we conclude from \eref{p3} that $\abs{\muf'(\slma)} = \mathcal{O}(1/\alpha^2)$. Substituting these limiting expressions into \eref{cos2_general} then gives us $\lim_{\alpha \rightarrow \infty} \rho(\alpha) = 1$, and this completes the proof.

\begin{remark}
When the set $\Lambda$ consists of a single element, which is the case for many signal acquisition models we have studied, $\lm_{c,\min} = \lm_{c, \max}$ and thus $\alpha_{c, \min} = \alpha_{c, \max}$. There then exists a single critical sampling ratio $\alpha_c$ separating the uncorrelated phase from the correlated one. For $\alpha < \alpha_c$, the estimates from the spectral method is asymptotically orthogonal to $\vxi_n$; for $\alpha > \alpha_c$, the estimates will be concentrated on the surface of a right-circular cone whose generating lines make an angle $\theta = \arccos(\sqrt{\rho(\alpha)})$ to the target vector $\vxi_n$. The situation is more complicated when $\Lambda$ contains multiple zero-crossings, in which case a finite number of correlated and uncorrelated phases can alternatively take place between $\alpha_{c, \min}$ and $\alpha_{c, \max}$. A concrete example demonstrating this situation is shown in the next subsection.
\end{remark}

\subsection{Multiple Phase Transitions: an Example}
\label{sec:counter_example}

Consider the following model:
\[
z_i = y_i = \begin{cases}
1, &\text{if } \abs{\va_i^\top \vxi_n} \in \mathcal{I}_1\\
\theta, &\text{if } \abs{\va_i^\top \vxi_n} \in \mathcal{I}_2\\
0, &\text{otherwise},
\end{cases}
\]
where $0 < \theta < 1$, and $\mathcal{I}_1, \mathcal{I}_2$ are two nonoverlapping intervals on the positive real axis. We also set $\nrm = \norm{\vxi_n} = 1$, and thus $\va_i^\top \vxi_n$ has the same distribution as a standard normal random variable, denoted by $s$. Define
\[
\beta_\ell = 2 \, \mathbb{E} \, \charfn_{\mathcal{I}_\ell}(s) \quad \text{and} \quad \omega_\ell = 2 \, \mathbb{E} \, (s^2 \charfn_{\mathcal{I}_\ell}(s)),
\]
where $\charfn_{\mathcal{I}_\ell}(\cdot)$ is the indicator function of $\mathcal{I}_\ell$, for $\ell = 1, 2$. As $z$ takes only $3$ different values, we can explicitly compute $\Delta(\lm)$ in \eref{delta_func} as
\[
\Delta(\lm) = \frac{\beta_1 \lm}{(\lm - 1)^2} + \frac{\theta \beta_2 \lm}{(\lm - \theta)^2} - \frac{\omega_1}{\lm - 1} - \frac{\theta \omega_2}{\lm - \theta}.
\]

Choose $\theta = 0.48$, $\mathcal{I}_1 = [4.7947,  4.9847]$, and $\mathcal{I}_2 = [0.8995,  0.8998]$. We then have $\beta_1 = 1.0086\times10^{-6}$, $\beta_2 = 1.5970\times 10^{-4}$, $\omega_1 = 2.3976\times 10^{-5}$ and $\omega_2 = 1.2926\times 10^{-4}$. In this case, $\Delta(\lm)$ turns out to have three zero-crossings:
\[
\lm_{c, 1} = 1.0765, \quad \lm_{c, 2} = 1.1844, \quad \lm_{c, 3} = 3.3127.
\]
By the mapping in \eref{blm_alpha}, they correspond to three critical sampling ratios:
\[
\alpha_{c, 1} = 3.6279 \times 10^3, \alpha_{c, 2} = 9.6302 \times 10^3, \alpha_{c, 3} = 2.0947 \times 10^5.
\]

\begin{figure}[t]
	\centering
	\includegraphics[width=0.9\linewidth]{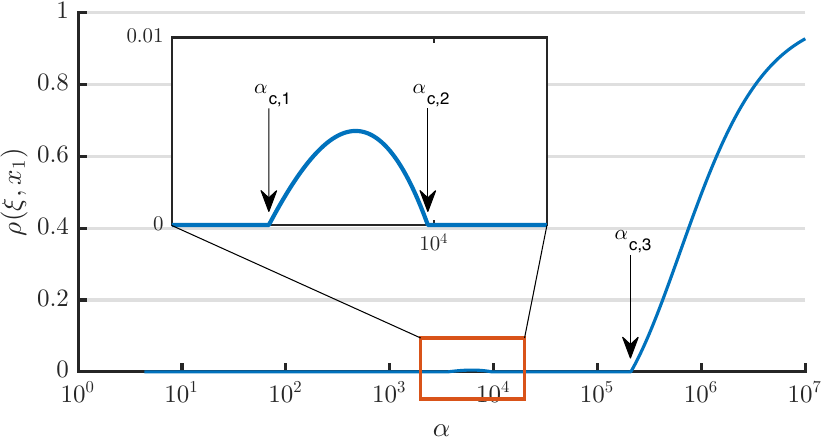}
	\caption{An example of multiple phase transitions. Shown in the figure is the limiting squared cosine similarity $\rho(\alpha)$ as a function of the sampling ratio $\alpha$. As $\alpha$ increases, the function $\rho(\alpha)$ alternates between the uncorrelated and correlated phases, with three phase transition points $\set{\alpha_{c, i}}_{1 \le i \le 3}$. The insert shows a zoomed-in view of the first correlated phase which takes place within the interval $(\alpha_{c,1}, \alpha_{c, 2})$.}\label{fig:counter_example}
\end{figure}

Using the characterization given in Theorem~\ref{thm:cos2}, we obtain the limiting values of the squared cosine similarity as a function of the sampling ratio $\alpha$. \fref{counter_example} illustrates this function $\rho(\alpha)$. We can see that, when $\alpha < \alpha_{c, 1}$, the estimates given by the spectral method are asymptotically uncorrelated with $\vxi_n$. When $\alpha$ is in the interval $(\alpha_{c, 1}, \alpha_{c, 2})$, however, the function $\rho(\alpha)$ has  a small ``bump'' (see the insert for a zoomed-in view), meaning that the estimates become asymptotically correlated with $\vxi_n$. However, the correlation returns to zero as $\alpha$ moves past the second phase transition point $\alpha_{c, 2}$. Finally, when $\alpha > \alpha_{c, 3}$, the estimates become correlated with $\vxi_n$ again, and $\rho(\alpha)$ tends to one as $\alpha \rightarrow \infty$. 

\begin{remark}
It would be desirable to obtain a deeper understanding of the above phenomenon involving multiple phase transitions. The example provided here is purely theoretical, as its phase transitions take place at very large values of $\alpha$. It will be interesting to explore other possible examples of multiple phase transitions with more practical values of $\alpha$. Moreover, as most signal acquisition models we have studied seem to involve only a single phase transition point, it will be interesting to seek easy-to-verify conditions for the function $\Delta(\lm)$ defined in \eref{delta_func} to have only one zero-crossing. We leave these as interesting open questions.
\end{remark}

\section{Discussion}
\label{sec:conclusion}

In this paper, we have presented a precise asymptotic characterization of the performance of a spectral method for estimating signals from generalized linear measurements with Gaussian sensing vectors. Our analysis also reveals a phase transition phenomenon that takes place at certain critical sampling ratios. Below a minimum threshold, estimates given by the methods are nearly orthogonal to the true signal $\vxi$, thus carrying no information; above a maximum threshold, the estimates become increasingly aligned with $\vxi$. The computational complexity of the spectral method is also markedly different in the two phases. Within the uncorrelated phase, the gap between the top two leading eigenvalues diminishes to zero. In contrast, a nonzero spectral gap emerges within the correlated phase. In this section, we close the paper by discussing some possible directions for extending and improving our results as well as their connections to related work in the literature.

\emph{The rate of convergence and more refined analysis.} The performance of the spectral method was first studied in \cite{Netrapalli:2013qv} for the problem of phase retrieval. In that paper, it is shown that, for each $\delta \in (0, 1)$, there is a constant $c_1(\delta)$ such that  $\rho(\xi_n, x_1^n) > 1 - \delta$ with high probability when
\[
m > c_1(\delta) n \log^3 n.
\]
This estimate of the sample complexity was improved to $m > c_2(\delta) n \log n$ in \cite{Candes:2013xy} and to $m > c_3(\delta) n$ in \cite{Chen:2015eu}. The key technical tools underlying these previous estimates are matrix concentration inequalities (see, \emph{e.g.}, \cite{Vershynin:12}), which guarantee that the spectral norm of the difference between the data matrix $\mD_m$ and its expectation $\mathbb{E} \mD_m$ will be small when the sampling ratio $m/n$ is sufficiently large. The closeness of the corresponding leading eigenvectors of $\mD_m$ and $\mathbb{E} \mD_m$ then follow from standard perturbation arguments. (See also our discussions towards the end of \sref{assumptions}.) Our work differs from and complements these finite-sample bounds in that we obtain sharp asymptotics to characterize the exact performance of the spectral method in the high-dimensional regime. A (theoretical) limitation of our analysis is that it is asymptotic in nature, requiring both $m, n \rightarrow \infty$. Although numerical simulations shown in \sref{numerical} indicate that the asymptotic predictions are accurate even for moderate signal dimensions, it will be useful to quantify the rate of convergence towards the asymptotic limits in future work.

Another possible direction to further refine our analysis is to consider second-order asymptotics at the level of central limit theorems (CLTs). See for instance \cite{Bai:2008} for a related CLT analysis for the extreme eigenvalues of spiked covariance models. 

\emph{Alternative initialization schemes.} The spectral method considered in this paper is certainly not the only choice for initialization purposes. For example, an interesting alternative is the simple linear estimator studied in \cite{Plan:2017}:
\begin{equation}\label{eq:linear_estimator}
\vx^n_\text{linear} = \frac{1}{m} \sum_{i=1}^m \mathcal{T}(y_i) \va_i.
\end{equation}
By using the moment calculations in \cite[Proposition 1.1]{Plan:2017} and bounding high-order moments, one can easily obtain that
\begin{equation}\label{eq:linear_asymp}
\rho(\vxi_n, \vx^n_\text{linear}) \cip \frac{(\mathbb{E} zs)^2}{(\mathbb{E} zs)^2 + \mathbb{E}z^2 / \alpha},
\end{equation}
where $s$ and $z$ are the random variables defined in \eref{sy}. 

Recall the function $g(s)$ introduced in \eref{g_func} and our discussions thereafter, where we point out that the spectral method is not suitable for acquisition models for which $g(s)$ is an odd function plus a constant. Such cases will pose no problem for the linear estimator in \eref{linear_estimator}. However, it is interesting to note that the linear estimator will be ineffective when $g(s)$ is an even function, as is the case in phase retrieval. To see this, we note that $\mathbb{E} zs = \mathbb{E} g(s)s = 0$ when $g(s)$ is even. It then follows from \eref{linear_asymp} that the linear estimator will be asymptotically uncorrelated with the target signal $\vxi_n$. 

For cases where the function $g(s)$ is neither odd nor even, the choice between the spectral method and the linear estimator is not as clear-cut. The spectral method exhibits phase transition behaviors with its estimates in the uncorrelated phase at small values of $\alpha$. In contrast, as shown in \eref{linear_asymp}, the performance of the linear estimator increases as a monotonic function of $\alpha$. As a result, in the regime of very small $\alpha$, the linear estimator will be preferable. For (moderately) larger values of $\alpha$, the comparison between the spectral method and the linear estimator cannot be easily made, as their performance also depends on the preprocessing function $\mathcal{T}(\cdot)$ used in \eref{D_mtx} and \eref{linear_estimator}. 

\emph{The incorporation of priors.} In this work, we assume that the target signal $\vxi_n$ is an arbitrary unknown (deterministic) signal. In many applications, the underlying signals satisfy additional constraints (such as sparsity). In \cite{Plan:2017}, the authors considered a two-step scheme, where the initial linear estimate given in \eref{linear_estimator} is further projected onto a set which encapsulates one's prior knowledge about $\vxi$. It will be interesting to consider and analyze similar projection schemes for the estimates obtained by the spectral method. 

\begin{figure}
	\centering
	\includegraphics[width=0.85\linewidth]{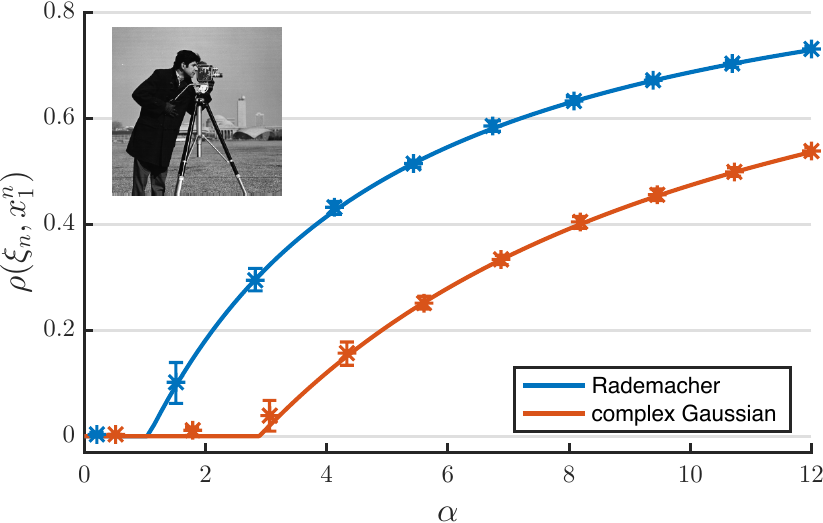}
	\caption{Demonstrating the potential \emph{universality} of our asymptotic characterizations. The figure shows the results of using the spectral method to estimate a $64 \times 64$ \emph{cameraman} image from phaseless measurements under Poisson noise. Theoretical predictions (the blue and red lines) are shown together with simulation results averaged over 16 independent trials. The error bars show one standard deviation. The blue line corresponds to real-valued sensing vectors drawn from the i.i.d. Rademacher ensemble, whereas the red line shows the results of using complex-valued sensing vectors drawn from the complex Gaussian distribution.}
	\label{fig:universality}
\end{figure}

%\emph{The need for initialization?}

\emph{Universality and more realistic sensing vectors.} Our asymptotic analysis assumes that the sensing vectors are real-valued i.i.d. Gaussian random vectors. Numerical simulations seem to suggest that the theoretical predictions given in Theorem~\ref{thm:cos2} remain valid for more general random measurement ensembles and for complex-valued sensing vectors. To demonstrate this, we show in \fref{universality} the results of applying the spectral method to estimate a $64 \times 64$ \emph{cameraman} image from phaseless measurements under Poisson noise:
\begin{equation}\label{eq:phase_Poisson}
y_i \sim \text{Poisson}(\,\abs{\va_i^* \vxi}^2) \quad \text{and } z_i = \min\set{y_i, \ub},
\end{equation}
where the bound $\ub$ is set to 5 and $\norm{\vxi}$ is normalized to 1 in our simulations. Two measurement ensembles are considered: real-valued sensing vectors whose elements are independent Rademacher ($\pm 1$) random variables, and complex-valued sensing vectors with elements drawn from the complex Gaussian distribution $\mathcal{N}(0, \tfrac{1}{2}) + j \mathcal{N}(0, \tfrac{1}{2})$. We see from the figure that the theoretical predictions (the solid lines) have excellent agreement with simulation results for this moderately-sized problem, even though the sensing vectors can be non-Gaussian. Rigorously establishing the validity of our asymptotic predictions without the Gaussian assumption will be an important future work. Thanks to the deterministic characterization given in Proposition~\ref{prop:fixed_point}, this task boils down to showing that the result of Proposition~\ref{prop:spike_pop} still holds when the sensing matrix consists of i.i.d. entries drawn from more general distributions. A related but more ambitious line of work will be to characterize the performance of the spectral method for structured and more practical sensing ensembles such as the coded diffraction scheme for phase retrieval with random modulation patterns. 

%Interestingly, as observed numerically in \cite{MondelliM:17}, the asymptotic predictions made for i.i.d. Gaussian matrices seem to match the results obtained from the coded diffraction scheme, provided that one uses a special form of the pre-processing function $\mathcal{T}(\cdot)$.

\emph{Low-rank matrix recovery.} The spectral method studied in this paper belongs to a more general theme. Let $\mat{X}^\star \in \R^{p \times n}$ be a rank-$r$ matrix and $\set{\mA_i}_{1 \le i \le m}$ a collection of sensing matrices of the same size as $\mat{X}^\star$. To  recover $\mat{X}^\star$ from linear measurements of the form $\set{y_i = \Tr \mA_i^\top \mat{X}}_i$, we can consider the following rank-constrained least squares problem
\[
\underset{\mat{X} \in \R^{p \times n}: \, \text{rank}\, \mat{X}=r}{\argmin} \sum_{i = 1}^m (y_i -  \Tr \mA_i^\top \mat{X})^2
\]
and try to solve it via projected gradient descent
\begin{equation}\label{eq:pgrad}
\mat{X}_{k+1} = \mathcal{P}_r \Big(\mat{X}_k + \mu \sum_{i = 1}^m (y_i - \Tr \mA_i^\top \mat{X}_k) \mA_i\Big),
\end{equation}
where $\mathcal{P}_r$ denotes projection onto the set of rank-$r$ matrices, and $\mu > 0$ is the step size. As pointed out in \cite{Tu:2015}, the spectral method studied in this paper can be viewed as the very first iteration of \eref{pgrad}, if we start the algorithm from $\mat{X}_0 = \vzero_{n \times n}$ and consider the special case of recovering a symmetric rank-one matrix (\emph{i.e.}, $r = 1$, $p = n$) with $\mA_i = \va_i \va_i^\top$. Thus, an interesting line of future research is to extend the results of this work, notably the key characterization given in Proposition~\ref{prop:fixed_point}, to more general settings with $r > 1$ and $p \neq n$ and to other sensing matrices. Such extensions will be useful in applications such as low-rank matrix recovery, covariance estimation, and blind deconvolution.

\appendix

\section*{}

\subsection{Sufficient Conditions for Assumption~\ref{a:infty} to Hold}
\label{appendix:infty}

In this appendix, we provide two sufficient conditions for Assumption~\ref{a:infty} to hold.

\emph{Case 1}: Suppose that the probability law of the random variable $z$ contains a point mass $c \, \delta(z - \ub)$ at its upper boundary $\ub$, where $c$ is some positive constant. This applies to the logistic regression model in Example~\ref{ex:logistic}, the subset algorithm \eref{subset} in Example~\ref{ex:phase}, the noisy phase retrieval model in \eref{phase_Poisson}, and the quantization model described in \sref{counter_example}.

In this case,
\[
\mathbb{E} \frac{z}{(\lm - z)^2} \ge \frac{c \ub}{(\lm - \ub)^2} \rightarrow \infty
\]
as $\lm \rightarrow \ub^+$. To verify the second expression in \eref{ub_infty}, let $h(z) = \mathbb{E}_{s \vert z}(s^2 \vert z)$. Since $\mathbb{P}(z = \ub) > 0$, we must have $h(\ub) > 0$. Thus,
\[
\mathbb{E} \frac{zs^2}{\lm - z} = \mathbb{E} \frac{z h(z)}{\lm - z} \ge \frac{c \ub h(\ub)}{\lm - \ub},
\]
which tends to $\infty$ as $\lm$ approaches $\ub$ from the right. 

\emph{Case 2}: Suppose that there exist some positive constants $c$ and $\varepsilon$ such that the probability density function $p_Z(z)$ of $z$ and the conditional moment $h(z)$ are both bounded below by $c$ for all $z \in [\ub - \varepsilon, \ub]$. The model in \eref{trimming} represents one such case. Under this setting,
\[
\begin{aligned}
\mathbb{E}\frac{zs^2}{\lm - z} &\ge \int_{\ub - \varepsilon}^{\ub} \frac{z h(z)}{\lm - z} p_Z(z) \dif z\\
 			&\ge (\ub - \varepsilon) c^2 \log\big(1 + \frac{\varepsilon}{\lm - \ub}\big) \xrightarrow{\lm \rightarrow \ub^+}\infty.
\end{aligned}
\]
Similarly, we can verify that $\mathbb{E} \frac{z}{(\lm - z)^2} \rightarrow \infty$ as $\lm \rightarrow \ub^+$.

\subsection{Norm Estimation}
\label{appendix:norm}

The spectral initialization method estimates the orientation of the vector $\vxi_n$ but it provides no information about its norm, as the eigenvector $\vx_1^n$ is always normalized. In many cases where the sensing vectors come from certain random ensembles, the norm $\norm{\vxi_n}$ can be accurately estimated from the measurements. 

As a simple illustrative example, we can consider the (noiseless) phase retrieval problem: $y_i = (\va_i^\top\vxi_n)^2$, where $\vxi_n$ is a deterministic unknown vector with $\nrm = \norm{\vxi_n}$, and the sensing vectors $\set{\va_i}$ are i.i.d. standard normal random vectors. Since $\va_i^\top\vxi_n \sim \mathcal{N}(0, \nrm^2)$, the measurement $y_i$ can be represented as
\[
y_i \sim \nrm^2 s_i^2,
\]
where $s_i$ (for $1 \le i \le m$) are i.i.d. standard normal random variables. A simple estimator of the norm is then
\begin{equation}\label{eq:mom_phase}
\widehat{\nrm} = \sqrt{\frac{\sum_{i=1}^m y_i}{m}},
\end{equation}
which is asymptotically consistent as $m \rightarrow \infty$.

More generally, consider an observation model $y_i \sim f(y \,\vert\, \va_i^\top \vxi_n)$, where $f(\cdot \,\vert\, \cdot)$ is a conditional probability density function and $\va_i \overset{\text{i.i.d.}}{\sim} \mathcal{N}(0, \mI_n)$. Again, writing $\va_i^\top \vxi_n = \nrm s_i$ for i.i.d. normal random variables $\set{s_i}$, we can represent the probability distributions of the measurements $\set{y_i}$ as
\[
y_i \overset{\text{i.i.d.}}{\sim} \int_{-\infty}^{\infty} f(y \,\vert\, \nrm s) \frac{1}{\sqrt{2\pi}} e^{-\frac{s^2}{2}} \, \dif s \bydef p_{\nrm}(y).
\] 

Let $w(\nrm) \bydef \mathbb{E}(y_i) = \int y  \, p_{\nrm}(y) \dif y$. If $w(\nrm)$ is monotonic on the positive real line, the \emph{method of moments} gives an estimator
\begin{equation}\label{eq:mom}
\widehat{\nrm}_{\text{MoM}} = w^{-1}\Big(\sum_{i=1}^m y_i / {m}\Big).
\end{equation}
We note that the estimator in \eref{mom_phase} is a special case of \eref{mom}. More generally, one could also estimate $\nrm$ by using maximum likelihood
\[
\widehat{\nrm}_{\text{MLE}} = \underset{\nrm > 0}{\arg\,\max} \sum_{i \le m} \log \, p_\nrm(y_i),
\]
whose asymptotic consistency can be established under standard conditions \cite{Lehmann:2003} on the parametric density function $p_\nrm(y)$.

\subsection{Proof of Proposition~\ref{prop:fixed_point}}
\label{appendix:fixed_point}

By a suitable choice of a transformation matrix
\[
\widetilde{\mW} = \pmatnocross \begin{pmat}[{|}]
1 & \vzero \cr \-
\vzero & {\mW} \cr
\end{pmat},
\]
where ${\mW} \in \R^{(n-1)\times(n-1)}$ is an orthogonal matrix involving the last $(n-1)$ rows and columns only, we can get a matrix
\begin{equation}\label{eq:D_tilde}
\widetilde{\mD} = \widetilde{\mW}^\top \mD \widetilde{\mW} = \pmatnocross\begin{pmat}[{| |}]
a & \widetilde{\vq}^\top & \vzero \cr \-
\widetilde{\vq} & \diag\set{\lambda_i^{\mP}}_{i \in \mathcal{I}_1} & \vzero \cr \-
\vzero & \vzero & \diag\set{\lambda_i^{\mP}}_{i \in \mathcal{I}_2}  \cr
\end{pmat},
\end{equation}
where $\mathcal{I}_1, \mathcal{I}_2$ are the two sets of indices defined in \eref{subsets} and $\widetilde{\vq}$ is a vector consisting of all the nonzero elements of $\mW^\top \vq$. Let $\lambda_1^{\widetilde{\mD}}$ and $\widetilde{\vx}_1$ be the largest eigenvalue of $\widetilde{\mD}$ and an associated unit-norm eigenvector, respectively. Clearly, $\lambda_1^{\mD} = \lambda_1^{\widetilde{\mD}}$ and $(\ve_1^\top \vx_1)^2 = (\ve_1^\top \widetilde{\vx}_1)^2$. Thus, we just need to consider $\widetilde{\mD}$ in our proof. 

Due to its block-diagonal form, the eigenvalues of $\widetilde{\mD}$ is the union of those of its top-left submatrix
\[
\mS = \pmatnocross
\begin{pmat}[{|}]
a 	& \widetilde{\vq}^\top \cr\-
\widetilde{\vq}	& \diag\set{\lambda_i^{\mP}}_{i \in \mathcal{I}_1} \cr
\end{pmat}.
\]
and those of its bottom-right submatrix $\diag\set{\lm_i^{\mP}: i \in \mathcal{I}_2}$. In particular,
\begin{equation}\label{eq:lms_v_lmp}
\lambda_1^{\widetilde{D}} = \lm_1^{\mS} \vee \max\set{\lm_i^{\mP}: i \in \mathcal{I}_2}.
\end{equation}

The eigenvectors associated with $\diag\set{\lm_i^{\mP}: i \in \mathcal{I}_2}$ are easy to characterize. Clearly, each $\lambda_i^{\mP}, i \in \mathcal{I}_2$ is an eigenvalue of $\widetilde{\mD}$, and it corresponds to an eigenvector $\ve_{j(i)}$, where $j(i) \ge 3$ is the row index of $\lambda_i^{\mP}$ in $\widetilde{\mD}$. 

The eigenvalues and eigenvectors of $\mS$ can also be precisely characterized. Due to its shape, $\mS$ is sometimes referred to in the literature as an arrowhead matrix \cite{OLeary:1990, Stor:2015}. It can be shown (see for instance \cite{Wilkinson:1988}[pp. 94 -- 97]) that $\lambda_1^{\mS}$ is the unique point within the interval $\lambda > \max\set{\lambda_i: i \in \mathcal{I}_1}$ to satisfy the equation
\begin{equation}\label{eq:lm_s}
a = \lambda_1^{\mS} + R(\lambda_1^{\mS}),
\end{equation}
where $R(\lambda)$ is the function defined in \eref{R_func}. (Alternatively, we can use the Laplace expansion to explicitly derive the characteristic polynomial of $\mS$ as
\[
(\lm - a) \prod_{i \in \mathcal{I}_1} (\lm - \lm_i^{\mP}) - \sum_{i \in \mathcal{I}_1} (\vw_i^\top \vq)^2 \prod_{j \in \mathcal{I}_1 \setminus i}(\lm - \lm^{\mP}_j).
\]
Then, by following similar arguments as those used in the proof of Lemma~\ref{lemma:L_rank_one}, we can reach the characterization \eref{lm_s} about $\lm_1^{\mS}$.) Furthermore, let $\vx_1^{\mS}$ be a unit-norm eigenvector of $\widetilde{\mD}$ associated with $\lambda_1^{\mS}$. It is easily checked that
\begin{equation}\label{eq:x1s_y}
\vx_1^{\mS} = \begin{bmatrix}1 & \vy & \vzero_r\end{bmatrix} / (1 + \norm{\vy}^2)^{1/2},
\end{equation}
where $\vy = (\lm_1^{\mS}\mI - \diag\set{\lm_i^{\mP}}_{i \in \mathcal{I}_1})^{-1} \widetilde{\vq}$ and $\vzero_r$ is a row vector of $r$ zeroes with $r$ being the cardinality of $\mathcal{I}_2$. It follows that
\begin{align}
(\ve_1^\top \vx_1^{\mS})^2 &= \Big(1 + \widetilde{\vq}^\top\big(\lm_1^{\mS}\mI - \diag\set{\lm_i^{\mP}}_{i \in \mathcal{I}_1}\big)^{-2} \widetilde{\vq}\Big)^{-1}\nonumber\\
	&= \big(1 + \vq^\top (\lm_1^{\mS}\mI - \mP)^{-2} \vq\big)^{-1}\nonumber\\
	&= \big(1 + R'(\lm_1^{\mS})\big)^{-1},\label{eq:x1s}
\end{align}
where $R'(\lm)$ denotes the derivative of the function $R(\lm)$.

To show the claim of the proposition, we consider the following three cases.

\emph{Case 1}: $\lm_1^{\mS} > \max\set{\lm_i^{\mP}: i \in \mathcal{I}_2}$. We choose $\mu^\ast = -1/R(\lm_1^{\mS})$, and thus $\lm_1^{\mS} =  R^{-1}(-1/\mu^\ast)$. It follows from Lemma~\ref{lemma:L_rank_one} that
\[
L(\mu^\ast) = \lm_1^{\mS} \vee \lm_1^{\mP} = \lm_1^{\mS} = \lm_1^{\widetilde{\mD}},
\]
where the second equality is due to the fact that
\begin{equation}\label{eq:lms_i1}
\lm_1^{\mS} > \max\set{\lm_i^{\mP}: i \in \mathcal{I}_1},
\end{equation}
and the last equality comes from \eref{lms_v_lmp}.

Using the identity \eref{lm_s} for $\lm_1^{\mS}$, we can also verify that $\mu^\ast$ indeed satisfies the equation \eref{mu_fixed_point}. (Its uniqueness is always guaranteed; see Remark~\ref{rem:mu} at the end of \sref{deterministic}.) The unit-norm leading eigenvector of $\widetilde{\mD}$ in this case is the vector $\vx_1^{\mS}$ defined in \eref{x1s_y}. Since $L(\mu) = R^{-1}(-1/\mu)$ in a neighborhood of $\mu^\ast$, the function $L(\mu)$ is differentiable at $\mu^\ast$ and
\begin{equation}\label{eq:dLdR}
L'(\mu^\ast) = \left(1 / R'(\lm_1^{\mS})\right) (\mu^\ast)^{-2}.
\end{equation}
Substituting \eref{dLdR} into \eref{x1s} leads to \eref{characterization}.

\emph{Case 2}: $\lm_1^{\mS} < \max\set{\lm_i^{\mP}: i \in \mathcal{I}_2}$, in which case $\lm_1^{\widetilde{D}} = \max\set{\lm_i^{\mP}: i \in \mathcal{I}_2} = \lm_1^{\mP}$, where the last equality is due to \eref{lms_i1}. The corresponding leading eigenvector has nonzero elements only in its last $r$ entries, where $r$ is the cardinality of $\mathcal{I}_2$. Thus, 
\begin{equation}\label{eq:cos2_0}
(\ve_1^\top \widetilde{\vx}_1)^2 = 0.
\end{equation}
We set $\mu^\ast = (\lm_1^{\mP} - a)^{-1}$. (Note that we are guaranteed to have $\mu^\ast > 0$. This can be verified by observing that $\lm_1^{\mP} > \lm_1^{\mS} = a - R(\lm_1^{\mS}) > a$, where the equality is due to \eref{lm_s}  and the last inequality follows from the fact that $R(\lm) < 0$.) Since $R(\lm)$ is a strictly increasing function, we have
\[
R^{-1}(-1/\mu^\ast) < R^{-1}(a - \lm_1^{\mS}) = \lm_1^{\mS} < \lm_1^{\mP},
\]
where the equality comes from \eref{lm_s}. It then follows from Lemma~\ref{lemma:L_rank_one} that $L(\mu^\ast) = \lm_1^{\mP} = \lm_1^{\widetilde{\mD}}$ and, moreover, $\mu^\ast$ satisfies the equation \eref{mu_fixed_point}.

To characterize the eigenvector, we note that $L(\mu) \equiv \lm_1^{\mP}$ in a neighborhood of $\mu^\ast$. We then have $L'(\mu^\ast) = 0$, which, together with \eref{cos2_0}, leads to \eref{characterization}.

\emph{Case 3}: $\lm_1^{\mS} = \max\set{\lm_i^{\mP}: i \in \mathcal{I}_2}$. This is a special case, where the algebraic multiplicity of the leading eigenvalue $\lm_1^{\widetilde{\mD}} = \lm_1^{\mS} = \lm_1^{\mP}$ is greater than one. The leading eigenvectors are not unique, and they can be any vector in the form of
\[
c_1 \vx_1^{\mS} + c_2 \vv,
\]
where $\vx_1^{\mS}$ is the eigenvector defined in \eref{x1s_y} and $\vv$ is an eigenvector associated with $\max\set{\lm_i^{\mP}: i \in \mathcal{I}_2}$, and $c_1, c_2$ are two constants satisfying $c_1^2 + c_2^2 = 1$. Since $\ve_1^\top \vv = 0$, we have from \eref{x1s} that
\begin{equation}\label{eq:interval2}
(\ve_1^\top \widetilde{\vx}_1)^2 \in [0, \big(1 + R'(\lm_1^{\mS})\big)^{-1}].
\end{equation}
Same as what we did in Case 2, we set $\mu^\ast = (\lm_1^{\mP} - a)^{-1}$. Following the same arguments there, we can show that $L(\mu^\ast) = \lm_1^{\mP} = \lm_1^{\widetilde{\mD}}$ and $\mu^\ast$ satisfies the equation \eref{mu_fixed_point}. Moreover, we can see that $L(\mu) =  R^{-1}(-1/\mu)$ for $\mu > \mu^\ast$ and $L(\mu) \equiv \lm_1^{\mP}$ for $\mu < \mu^\ast$. The function $L(\mu)$ is not differentiable at $\mu^\ast$, but its right and left derivatives do exist. It is easy to get $\partial_{+}L(\mu^\ast) =  \left(1 / R'(\lm_1^{\mS})\right) (\mu^\ast)^{-2}$ [see \eref{dLdR}] and $\partial_{-}L(\mu^\ast) = 0$. Substituting these quantities into \eref{interval2}, we reach the characterization given in \eref{interval1}.

\subsection{Proof of Proposition~\ref{prop:bulk_spike}}
\label{appendix:bulk_spike}

To establish the almost-sure convergence of the random measure $f^{\mM_m}(\lambda)$ to the probability law of $z$, we just need to show that, almost surely, the empirical distribution function
\[
F^{\mM_m}(\lm) = \frac{1}{m-1} \#\set{2 \le j \le m: \lm_j^{\mM_m} \le \lm}
\]
converges to $F_z(\lm)$, the cumulative distribution function of z, at all points $\lm$ where $F_z(\lm)$ is continuous. Since $\mM_m$ is a rank-one perturbation of the diagonal matrix $\mZ$, standard interlacing theorems (see \cite[Theorem 4.3.4]{HornJ:85}) give us
\begin{equation}\label{eq:interlacing_rank_one}
\lm_k^{\mZ} \ge \lm_{k+1}^{\mM_m} \ge \lm_{k+2}^{\mZ},
\end{equation}
for $1 \le k \le m-2$. Let $F^{\mZ}(\lm) = \frac{1}{m} \#\set{1 \le j \le m: z_j \le \lm}$ be the empirical distribution function of the eigenvalues of $\mZ$. We can then easily verify from \eref{interlacing_rank_one} that
\begin{equation}\label{eq:Fm_Fz}
m F^{\mZ}(\lm) - 2 \le (m-1) F^{\mM_m}(\lm) \le m F^{\mZ}(\lm) + 1.
\end{equation}
Since $\set{z_i}_{1 \le i \le m}$ is an i.i.d. sample of the random variable $z$, with probability one $F^{\mZ}(\lm)$ converges to $F_z(\lm)$ all all points $\lm$ where $F_z(\lm)$ is continuous. It then follows from \eref{Fm_Fz} that $F^{\mM_m}(\lm)$ converges almost surely to the same limit $F_z(\lm)$.

To study the leading eigenvalue $\lm_1^{\mM_m}$, we use Lemma~\ref{lemma:L_rank_one}. To apply that result, we require $\vv = [z_1 s_1, z_2 s_2, \ldots, z_m s_m]^\top$ as defined in \eref{q} to be not equal to the all-zero vector. This condition holds almost surely for all sufficiently large $m$. To see this, we note that $\mathbb{P}(z_i = 0) < 1$, as otherwise assumption~\ref{a:pos_corr} will not hold. Moreover, $s_i \neq 0$ with probability one. It follows that the i.i.d. sequence $z_1 s_1, z_2 s_2, z_3 s_3, \ldots$ has an infinite number of nonzero elements. Thus, almost surely, the $m$-dimensional vector $\vv \neq \vzero$ for sufficiently large $m$.

Applying \eref{L_rank_one} to our case, we have $\lm_1^{\mM_m} = R_m^{-1}(-1/\mu) \vee \max\set{z_i}_{1 \le i \le m}$, where
\[
R_m(\lm) = \frac{1}{m}\sum_{i=1}^{m} \frac{z_i^2 s_i^2}{z_i - \lm}
\]
with this function defined on $\lm > \max\set{z_i}_{1 \le i \le m}$. Since $R_m^{-1}(-1/\mu) > \max\set{z_i}_{1 \le i \le m}$, we can further simplify the characterization to
\[
\lm_1^{\mM_m} = R_m^{-1}(-1/\mu).
\]
For every $\lm > \ub$, with $\ub$ being the upper bound of the support of the probability distribution of $z$, it follows from the strong law of large numbers that $R_m(\lm)$ converges almost surely to
\[
\mathbb{E}\frac{z^2 s^2}{z - \lm} = -Q(\lm),
\]
where $Q(\lm)$ is defined in \eref{q_lm}. On its domain $\lm > \ub$, the function $-Q(\lm)$ is strictly increasing and thus it admits a functional inverse $(-Q)^{-1}(x) = Q^{-1}(-x)$. Applying Lemma~\ref{lemma:inverse} in Appendix~\ref{appendix:auxiliary}, we have
\[
\lm_1^{\mM_m} = R_m^{-1}(-1/\mu) \asc Q^{-1}(1/\mu)
\]
as $m \rightarrow \infty$.

\subsection{Auxiliary Lemmas}
\label{appendix:auxiliary}

We prove here two auxiliary lemmas that are used in our proofs of Proposition~\ref{prop:bulk_spike} and Theorem~\ref{thm:cos2}.

\begin{lemma}\label{lemma:inverse}
Let $\set{f_n(x)}_{n \ge 1}$ be a family of (random) functions defined on an open interval $(a, b)$. Each $f_n(x)$ is continuous and nondecreasing. For each $x \in (a, b)$, $f_n(x) \asc f(x)$ as $n \rightarrow \infty$, where $f(x)$ is a continuous and nondecreasing function. Then, for any sequence $\set{x_n} \subset (a, b)$ with $x_n \asc x^\ast \in (a, b)$, we have
\begin{equation}\label{eq:fnxn}
f_n(x_n) \asc f(x^\ast).
\end{equation}
If, in addition, the functions $\set{f_n(x)}$ and $f(x)$ are strictly increasing, we denote by $\set{f_n^{-1}(x)}_{n \ge 1}$ and $f^{-1}(x)$ the corresponding functional inverses. Assume that the domains of $\set{f_n^{-1}(x)}_{n \ge 1}$ and $f^{-1}(x)$ contain a common open interval $\mathcal{I}$. Then for any sequence $\set{y_n}_{n \ge 1} \subset \mathcal{I}$ such that $y_n \asc y \in \mathcal{I}$, we have
\begin{equation}\label{eq:fn_inverse}
f_n^{-1}(y_n) \asc f^{-1}(y).
\end{equation}
\end{lemma}
\begin{IEEEproof}
We first show \eref{fnxn}. Let $\beta_k = x^\ast - h / k$ for $k = 1, 2, \ldots$ be a sequence that converges to $x^\ast$ from the left. We choose $h <  \min\set{x^\ast - a, b - x^\ast}$ so that the entire sequence stays within the interval $(a, b)$. Similarly, define a sequence $\gamma_k = x^\ast + h/k$, for $k = 1, 2, \ldots$, that converges to $x^\ast$ from the right. Denote by $\mathcal{A}$ the intersection of the event that $f_n(x) \rightarrow f(x)$ for all $x \in \set{\beta_k}_k \cup \set{\gamma_k}_k$ and the event that $x_n \rightarrow x$. Clearly, $\mathbb{P}(\mathcal{A}) = 1$. Next, we show that \eref{fnxn} holds within this almost sure event.

Fix $k \ge 1$. As $x_n \rightarrow x^\ast$, we have $\beta_k \le x_n \le \gamma_k$ for all sufficiently large $n$. By the monotonicity of $f_n(x)$, 
\[
f_n(\beta_k) \le f_n(x_n) \le f_n(\gamma_k).
\]
It follows that
\[
f(\beta_k) \le \underset{n}{\lim\inf} \,f_n(x_n) \le \underset{n}{\lim\sup} \, f_n(x_n) \le f(\gamma_k).
\]
As $k$ is arbitrary, we take the $k \rightarrow \infty$ limit, which leads to $\lim_n f_n(x_n) = f(x)$ by the continuity of $f(x)$.

The proof of \eref{fn_inverse} is similar. We establish it under the additional assumption that $\set{f_n(x)}$ and $f(x)$ are strictly increasing. Construct two sequences $\set{\beta_k}$ and $\set{\gamma_k}$ as above, with $x^\ast$ replaced by $f^{-1}(y)$. Also define the event $\mathcal{A}$ similarly. We show that, within the almost sure event $\mathcal{A}$, we have $f_n^{-1}(y_n) \rightarrow f^{-1}(y)$.

Fix $k \ge 1$. Since $f(x)$ is strictly increasing, $\beta_k < f^{-1}(y) < \gamma_k$ implies that
\[
f(\beta_k) < y < f(\gamma_k).
\]
As $f_n(\beta_k) \rightarrow f(\beta_k)$, $f_n(\gamma_k) \rightarrow f(\gamma_k)$ and $y_n \rightarrow y$, the inequalities
\[
f_n(\beta_k) < y_n < f_n(\gamma_k),
\]
hold for all sufficiently large $n$. By the strict monotonicity of $f_n(x)$,
\[
\beta_k < f_n^{-1}(y_n)  < \gamma_k,
\]
for all sufficiently large $n$. It then follows that $\beta_k \le \lim\inf_n f_n^{-1}(y_n) \le \lim\sup_n f_n^{-1}(y_n) \le \gamma_k$, for each $k$. As $\beta_k \rightarrow f^{-1}(y)$ and $\gamma_k \rightarrow f^{-1}(y)$, we are done.
\end{IEEEproof}

\begin{lemma}\label{lemma:derivative}
Let $\set{f_n(x)}_{n \ge 1}$ be a sequence of (random) convex functions defined on an open interval $(a, b)$. For each $x \in (a, b)$, $f_n(x) \asc f(x)$. Let $\set{x_n}_{n \ge 1} \subset (a, b)$ be a sequence such that $x_n \asc x^\ast$ for some $x^\ast \in (a, b)$. If $f(x)$ is differentiable at $x^\ast$, then
\begin{equation}\label{eq:lr_derivatives}
\partial_- f_n(x_n) \asc f'(x^\ast) \quad\text{and}\quad \partial_+ f_n(x_n) \asc f'(x^\ast),
\end{equation}
where $\partial_- f_n(x)$ and $\partial_+ f_n(x)$ denote the left and right derivatives of $f_n(x)$, respectively.
\end{lemma}
\begin{IEEEproof}
Similar to the proof of Lemma~\ref{lemma:inverse}, we construct two sequences: $\set{\beta_k}_{k \ge 1}$ is strictly increasing and converges to $x^\ast$ from the left, whereas $\set{\gamma_k}_{k \ge 1}$ is strictly decreasing and converges to $x^\ast$ from the right. Denote by $\mathcal{A}$ the intersection of the event that $f_n(x) \rightarrow f(x)$ for all $x \in \set{\beta_k}_k \cup \set{\gamma_k}_k$ and the event that $x_n \rightarrow x^\ast$. It is easily checked that $\mathbb{P}(\mathcal{A}) = 1$. Next, we establish \eref{lr_derivatives} within this almost sure event.

For any $i < j$, since $\beta_i < \beta_j < x^\ast$ and $x_n \rightarrow x^\ast$, we must have $\beta_i < \beta_j < x_n$ for all sufficiently large $n$. By the convexity of $f_n(x)$, its left derivatives always exist and we have
\[
\partial_- f_n(x_n) \ge \frac{f_n(\beta_i) - f_n(\beta_j)}{\beta_i - \beta_j},
\]
for all sufficiently large $n$. It follows that
\[
\underset{n}{\lim\inf} \; \partial_- f_n(x_n) \ge \frac{f(\beta_i) - f(\beta_j)}{\beta_i - \beta_j}
\]
for all $i < j$. Since
\[
\lim_{i \rightarrow \infty} \left(\lim_{j \rightarrow \infty} \frac{f(a_i) - f(a_j)}{a_i - a_j}\right) = f'(x^\ast),
\]
we must have
\begin{equation}\label{eq:liminf}
\underset{n}{\lim\inf} \; \partial_- f_n(x_n) \ge f'(x^\ast).
\end{equation}
Working with the sequence $\set{\gamma_k}_{k \ge 1}$ and using similar arguments as above, we can show that
\begin{equation}\label{eq:limsup}
\underset{n}{\lim\sup} \; \partial_+ f_n(x_n) \le f'(x^\ast).
\end{equation}
Since $\lim\sup_n \partial_-f_n(x_n) \le \lim\sup_n \partial_+f_n(x_n)$, we use \eref{liminf} and \eref{limsup} to conclude that $\lim_n \partial_-f_n(x_n)$ exists and that it is equal to $f'(x^\ast)$. By similar arguments, the same claim also holds for the sequence $\set{\partial_+f_n(x_n)}$, and thus the proof is complete.
\end{IEEEproof}

%\begin{itemize}
%\item Results obtained in the first paper that proposed spectral initialization
%
%\item Results in later improvements (Wirtinger flow or truncated Wirtinger Flow)
%
%\item Mention the more general idea of (nonconvex) projected gradient.
%
%\item Spiked covariance model
%
%\item Weyl purturbation eigenvalue, eigenvector
%
%\item Davis-Kahan
%
%\end{itemize}



%\bibliographystyle{IEEEtran}

\bibliographystyle{IEEEtran}
\bibliography{refs}

\end{document}